\documentclass[journal]{IEEEtran}
\RequirePackage{filecontents}

\usepackage{epsfig,amsmath,amssymb,epsf,amsthm,scalefnt,multirow,subfig}
\usepackage{xcolor}
\usepackage{float}
\usepackage{cite}
\usepackage{psfrag}\usepackage{graphics,amsmath,amssymb,graphicx,amsthm,scalefnt,multirow,dsfont,subfig}
\usepackage{algorithm}
\usepackage{algpseudocode}
\usepackage{verbatim}
\usepackage{tabularx,booktabs}
\usepackage{multicol}
\usepackage{stfloats}
\usepackage{bm}

\newtheorem*{lemma*}{Lemma}

 \def\cB{{\mathcal{B}}} \def\cC{{\mathcal{C}}}

 \def\cN{{\mathcal{N}}} \def\cO{{\mathcal{O}}} 
\def\cQ{{\mathcal{Q}}}   
\def\cU{{\mathcal{U}}}

\def\bPhi{{\boldsymbol{\Phi}}} \def\bphi{{\pmb{\phi}}}
\def\bGamma{{\pmb{\Gamma}}} \def\bgamma{{\pmb{\gamma}}}
\def\bOmega{{\pmb{\Omega}}}

\def\b0{{\pmb{0}}} 
\def\bpsi{{\pmb{\psi}}}
\def\bPi{{\pmb{\Pi}}}\def\bpi{{\pmb{\pi}}}

\def\ba{{\mathbf{a}}} \def\bb{{\mathbf{b}}}  
\def\bee{{\mathbf{e}}}  \def\bg{{\mathbf{g}}} \def\bh{{\mathbf{h}}}
   
 \def\bn{{\mathbf{n}}}  
   
\def\bu{{\mathbf{u}}} \def\bv{{\mathbf{v}}} \def\bw{{\mathbf{w}}} 
\def\by{{\mathbf{y}}}   

\def\bA{{\mathbf{A}}} \def\bB{{\mathbf{B}}}  \def\bD{{\mathbf{D}}}
   
\def\bI{{\mathbf{I}}}   
   
   \def\bT{{\mathbf{T}}}
  \def\bW{{\mathbf{W}}} 
\def\bY{{\mathbf{Y}}}

\newcommand{\C}{\mathbb{C}}

\theoremstyle{remark}

\begin{document}
	
	\title{Meta-Heuristic Fronthaul Bit Allocation for Cell-free Massive MIMO Systems}
	
	\author{Minje Kim,~\IEEEmembership{Graduate Student Member,~IEEE}, In-soo Kim,~\IEEEmembership{Member,~IEEE}, \\and~Junil Choi,~\IEEEmembership{Senior Member,~IEEE}
		\thanks{
			Minje Kim and Junil Choi are with the School of Electrical Engineering, Korea Advanced Institute of Science and Technology, Daejeon 34141, South Korea
			(e-mail: mjkim97@kaist.ac.kr; junil@kaist.ac.kr).
			
			In-soo Kim is with Wireless Research \& Development (WRD), Qualcomm Technologies Inc., San Diego, CA 92121 USA (e-mail: \mbox{insookim}@qti.qualcomm.com).
			
		}
	}
	\maketitle
	
	\begin{abstract}
		Limited capacity of fronthaul links in a cell-free massive multiple-input multiple-output (MIMO) system can cause quantization errors at a central processing unit (CPU) during data transmission, complicating the centralized rate optimization problem. Addressing this challenge, we propose a harmony search (HS)-based algorithm that renders the combinatorial non-convex problem tractable. One of the distinctive features of our algorithm is its hierarchical structure: it first allocates resources at the access point (AP) level and subsequently optimizes for user equipment (UE), ensuring a more efficient and structured approach to resource allocation. Our proposed algorithm deals with rigorous conditions, such as asymmetric fronthaul bit allocation and distinct quantization error levels at each AP, which were not considered in previous works. We derive a closed-form expression of signal-to-interference-plus-noise ratio (SINR), in which additive quantization noise model (AQNM) based distortion error is taken into account, to define the mathematical expression of spectral efficiency (SE) for each UE. Also, we provide analyses on computational complexity and convergence to investigate the practicality of proposed algorithm. By leveraging various performance metrics such as total SE and max-min fairness, we demonstrate that the proposed algorithm can adaptively optimize the fronthaul bit allocation depending on system requirements.
		Finally,  simulation results show that the proposed algorithm can achieve satisfactory performance while maintaining low computational complexity, as compared to the exhaustive search method.
	\end{abstract}

	\section{Introduction}\label{sec1}
	As the demand for unprecedented data rate, reliability, and connectivity increases in 5G and beyond era, many wireless communication technologies are studied to fulfill the requirements.
	Massive multiple-input multiple-output (MIMO) is expected to be a key technology for upcoming wireless communication systems, promising energy-efficient and high-throughput communications \cite{Gesbert:2007,Marzetta:2010,Larsson:2014,Boccardi:2014}.
	One potential implementation of massive MIMO systems is cell-free massive MIMO systems, which can jointly process all data from/to a large number of widely distributed access points (APs) \cite{Ngo:2017, Nayebi:2017, Chen:2018}. In cell-free massive MIMO systems, the absence of cell-concept and cooperative resource utilization can get rid of inter-cell interference that negatively affects the communication performances of cell-edge users.
	Therefore, cell-free massive MIMO systems can achieve enhanced fairness with more uniform coverage than conventional cellular systems \cite{Shidong:2004, Truong:2013}. To fully exploit the benefits of centralized processing in cell-free massive MIMO systems, reliable data transmission link between each AP and a central processing unit (CPU), called as a fronthaul link, is essential. In practice, however, utilizing perfect fronthaul links is infeasible \cite{limitedfronthaul1, limitedfronthaul3}, and the capacity of fronthaul link is a significant constraint that needs to be addressed in cell-free massive MIMO systems.

	One possible way to support the limited fronthaul capacity is to compress the data exchanged between the CPU and the APs by the quantizer. Recently, several works took the data quantization at the fronthaul into consideration in cell-free massive MIMO systems \cite{limitedfronthaul2,Kim:2022,Bashar:2019:maxminrate,Bashar:2019:energyefficiency,Liang:2015,Rajapaksha:2023}. 
	These works handled various objectives with the uniform quantization that is modeled based on the additive quantization noise model (AQNM).  
	In \cite{limitedfronthaul2}, the achievable rate of the limited fronthaul cell-free massive MIMO system was analyzed under the hardware impairment assumption. 
	The authors of \cite{Kim:2022} proposed the codebook, which minimizes the channel estimation error in the cell-free massive MIMO system when both the limited fronthaul capacity and the low-resolution analog-to-digital converters/digital-to-analog converters (ADCs/DACs) exist.
	These works, however, were based on the fixed fronthaul bits.
	In contrast, several other works considered fronthaul bit allocation techniques\cite{Bashar:2019:maxminrate, Bashar:2019:energyefficiency,Liang:2015,Rajapaksha:2023}.
	In particular, the cell-free massive MIMO systems were investigated in the aspects of the max-min rate in \cite{Bashar:2019:maxminrate} and the energy-efficiency in \cite{Bashar:2019:energyefficiency}, under the equal fronthaul bit allocation, which assigns the same amount of bits to all the APs.
	While the concept of independent bit allocation for each AP was introduced in \cite{Liang:2015}, it considered the ideal quantizer that provides the same quantization error regardless of the number of allocated bits. In \cite{Rajapaksha:2023}, the machine-learning based fronthaul capacity allocation was proposed, but it considered allocating continuous capacity that might not reflect the practical implementation.
	
	In this paper, we consider a cell-free massive MIMO system with limited fronthaul links where allocated bits can be independently optimized. Also, we consider distinct quantization errors for different amounts of allocated bits. Unfortunately, the fronthaul bit allocation problem consists of both continuous and discrete variables, making the objective formulation highly non-convex. Therefore, we propose a meta-heuristic fronthaul bit allocation algorithm to tackle the non-convex combinatorial problem. The meta-heuristic is one of the optimization techniques, which has the strength in solving combinatorial problems \cite{Blum:2001}. 
	In contrast to exact methods, the goal of the meta-heuristic is to find an acceptable solution with a significantly reduced complexity. There are various meta-heuristic algorithms, including ant-colony optimization (ACO) algorithm \cite{ACO}, genetic algorithm (GA) \cite{Generic}, particle swarm optimization (PSO) algorithm \cite{PSO}, and simulated annealing (SA) algorithm \cite{SA}. Among them, we develop a fronthaul bit allocation approach based on harmony search (HS) algorithm, which finds the near-optimal solution via population-based search where multiple initial points are generated and combined during the procedure \cite{Geem:2001}. 
	\begin{table*}[htbp]
		\centering
		\captionsetup{justification=centering, labelsep=newline, font={smaller,sc}}
		\caption{Feature Comparison of Meta-heuristic Algorithms}
		\label{tab:metaheuristic-comparison}
		\setlength{\tabcolsep}{2.5em} 
		\begin{tabularx}{0.83\linewidth}{@{}ccccc@{}}
			\toprule
			\textbf{Algorithm} & \textbf{Type} & \textbf{Variable} & \textbf{Target Problem} & \textbf{Used Solutions} \\
			\midrule
			ACO & Population & Continuous/Discrete & Shortest Path & Multiple \\
			GA & Population & Continuous/Discrete & Function Optimization & Multiple \\
			HS & Population & Continuous/Discrete & Function Optimization & Multiple \\
			PSO & Population & Continuous & Function Optimization & Multiple \\
			SA & Trajectory & Continuous/Discrete & Function Optimization & Single \\
			\bottomrule
		\end{tabularx}
	\end{table*}
	The comprehensive comparison outlined in Table I emphasizes the preference for the HS algorithm over alternative options, including ACO, GA, PSO, and SA. Remarkably, ACO primarily targets the shortest path predicaments, making it less suitable for function optimization tasks like fronthaul bit allocation. Conversely, GA, HS, and PSO are aligned with function optimization, rendering them more fitting choices. PSO, however, primarily functions with continuous variables, which may not align with the problem's discrete nature. In contrast, HS caters to both continuous and discrete variables, augmenting its adaptability. While integer PSO in \cite{integerPSO} can support discrete variables, approximation process for discretization possibly induces undesired loss. During update procedure, GA utilizes pairs of parents in the given population to generate new offsprings, which has the potential to result in overlapping solutions \cite{premature}. The HS algorithm, however, can avoid generating redundant solutions with its inherit nature.  
	 Therefore, we propose an algorithm based on HS, and its superiority over other algorithms will be demonstrated in Section~\ref{sec5}.

	The main contributions of this paper are as follows.
	\begin{itemize}
		\item Our innovative HS-based algorithm efficiently allocates limited fronthaul bit resources in a cell-free massive MIMO system. It follows a hierarchical structure, initially assigning bits to APs and subsequently reallocating them to user equipments (UEs). This flexible approach can allocate bits exclusively to APs, UEs, or both, accommodating scenarios with constrained computational resources.
		
		\item We derive a signal-to-interference-plus-noise ratio (SINR) expression that accounts for non-uniform data quantization errors, addressing a crucial aspect of our approach. To tackle the inherent non-convexity and discrete optimization variables in the bit allocation problem, we introduce our HS-based algorithm. This method effectively manages computational complexity while delivering solutions that closely approximate those obtained through exhaustive search, ensuring high-performance outcomes.
		
		\item Our proposed algorithm demonstrates remarkable adaptability, capable of accommodating various objective functions such as the maxmin fairness with minor modifications. Its versatility makes it a valuable tool for optimizing resource allocation in diverse scenarios.
	\end{itemize}
	
	The remainder of this paper is structured as follows. In Section~\ref{sec2}, we describe our cell-free massive MIMO system model with limited capacity fronthauls. In Section~\ref{sec3}, we analyze the SE in the cell-free massive MIMO system by deriving the SINR in closed-form. To address the bit allocation problem, a low-complexity HS-based algorithm is proposed in Section~\ref{sec4}. The performance of the proposed algorithm in terms of both the total SE and max-min fairness is demonstrated through simulation results in Section~\ref{sec5}. Finally, the conclusion follows in Section~\ref{conclusion}.
	
	In Sections \ref{sec2} and \ref{sec3} of this paper, we build upon the foundational derivations presented in \cite{limitedfronthaul2} to establish the framework for our system model and subsequent fronthaul bit allocation analysis. While the core derivations are drawn from \cite{limitedfronthaul2}, we have made specific modifications and adaptations to suit the objectives and context of our work. This approach allows us to provide a self-contained and comprehensive discussion of our contributions within the framework of our proposed solution.

	\textbf{Notation:} Scalars, column vectors, and matrices are represented by italic letters, lower boldface letters, and upper boldface letters, respectively. Superscripts $(\cdot)^*$, $(\cdot)^{-1}$, $(\cdot)^{\mathrm{T}}$,  and $(\cdot)^{\mathrm{H}}$ denote the conjugate, matrix inverse, matrix transpose, and matrix conjugate transpose, respectively. For any vector $\ba$, $\|\ba\|$ refers to its Euclidean norm, and $\mathrm{diag}(\ba)$ is used to denote the matrix whose diagonal elements consist of the elements in $\ba$. $\bI_m$ represents the $m \times m$ identity matrix. $\mathbb{C}^{m\times n}$ and $\mathbb{R}^{m\times n}$ are used for the sets of all $m \times n$ complex and real matrices, respectively. $\cC\cN(0,\sigma^2)$ refers to a circularly symmetric complex Gaussian distribution whose mean is zero and variance is $\sigma^2$.
	
	\begin{figure}[t]
		\centering
		\includegraphics[width=8 cm]{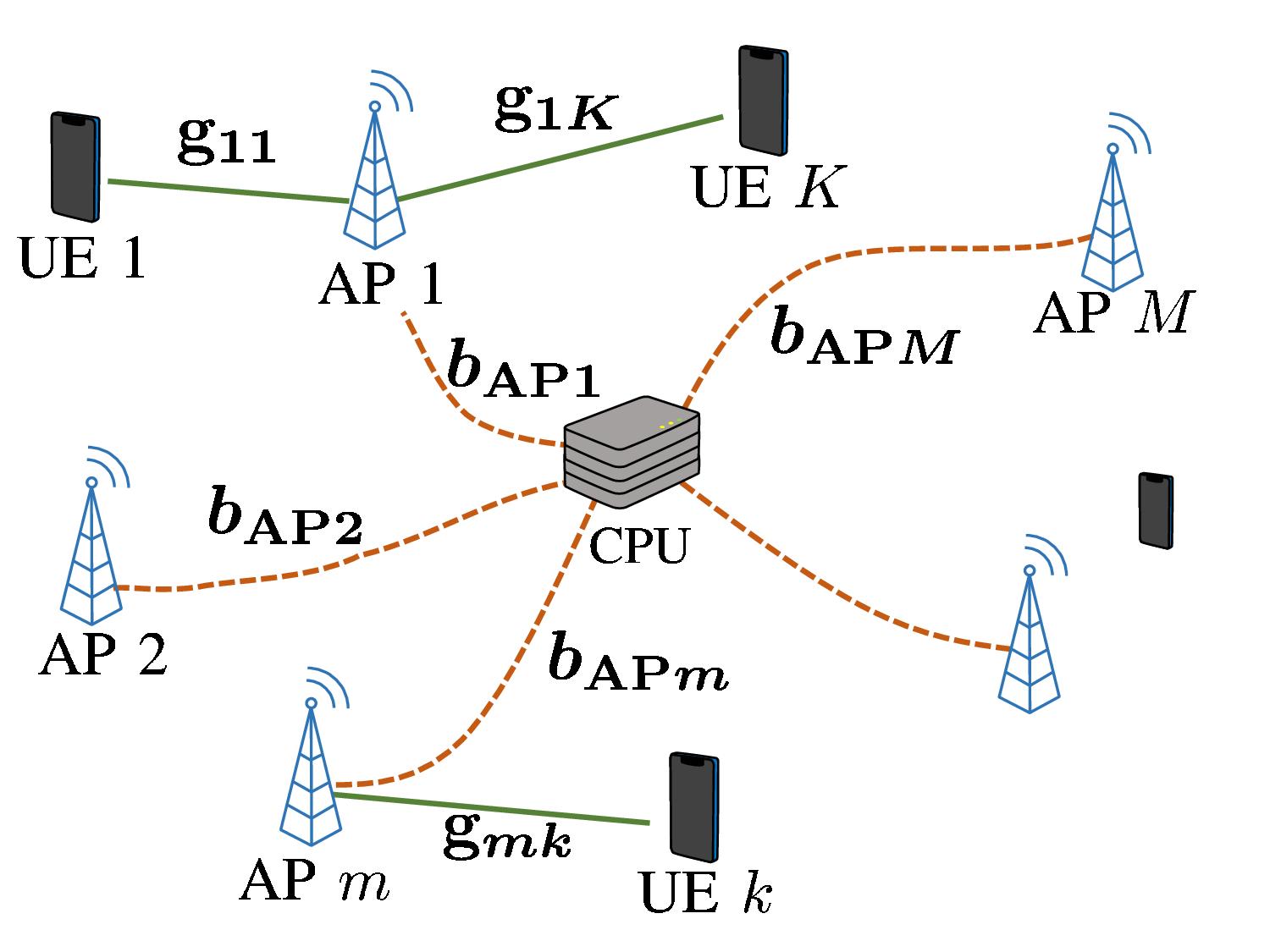}%
		\caption{System model.}
		\label{sysmodel}
	\end{figure}
	
	\section{System Model}\label{sec2}
	
	We consider a cell-free massive MIMO system with $M$ APs, each equipped with $N$ antennas, and $K$ single-antenna UEs. Each AP is connected to the CPU through the fronthaul link as described in Fig.~\ref{sysmodel}. In the considered scenario, $M$ APs simultaneously serve $K$ UEs using the common time and frequency resources.
	To consider practicality of cell-free massive MIMO systems, we assume non-ideal fronthaul links whose bits are constrained as
	\begin{align} \label{bit constaint}
	\sum_{m=1}^M\sum_{k=1}^K b_{mk} \le b_\mathrm{max},
	\end{align}
	where $b_\mathrm{max}$ denotes the maximum number of available fronthaul bits, and $b_{mk}$ refers to the allocated fronthaul bits for the $k$-th UE's signal at the fronthaul link between the $m$-th  AP and CPU.
	With this constraint, we investigate the operation of the considered system in the following subsections.
	
	In Section \ref{2-a}, we first explain the uplink channel estimation phase. Then, we elaborate the uplink data transmission phase in Section \ref{2-b}. We also discuss the quantization model due to the limited fronthaul capacity and the receiver filter design at the CPU in Sections \ref{2-c} and \ref{2-d}, respectively.
	
	\subsection{Uplink Channel Estimation} \label{2-a}
	In this subsection, we provide the channel model and how to estimate the channels at the APs during the uplink channel estimation phase.
	Let  ${\bg_{mk}\in \C^{N \times 1}}$ be the channel between the $k$-th UE and the $m$-th AP, which is modeled as
	\begin{align}
		\label{channel}
		\bg_{mk}=\sqrt{\beta_{mk}} \bh_{mk},
	\end{align}
	where $\beta_{mk}$ is the large-scale fading, and ${\bh_{mk} \sim \cC\cN(0,\bI_N)}$ denotes the small-scale fading. We assume the large-scale fading $\beta_{mk}$ varies considerably slower than the small-scale fading  $\bh_{mk}$ such that the AP can obtain the long-term channel information $\beta_{mk}$.
	Also, the channel $\bg_{mk}$ in \eqref{channel} is assumed to be independent for all $m$ and $k$.


	
	
	During the uplink channel estimation phase, all the UEs transmit pilot sequences, and the APs estimate the channels.
	To prevent pilot contamination, all UEs transmit their own orthogonal pilot sequences $\sqrt{\tau_p}\bpsi_k \in \C^{\tau_p \times 1} $ that satisfy $\lvert \bpsi_k^\mathrm{H}\bpsi_{k'} \rvert = \delta_{kk'} $, where $\tau_p$ denotes the length of the pilot sequence, and $\delta_{kk'}$ refers to the Kronecker delta function.\footnote{When considering non-orthogonal pilot sequences, the main difference from the orthogonal case is the added interference term due to the pilot contamination. This trait ensures our algorithm can be easily modified for non-orthogonal scenarios. To highlight the flexibility and performance of our algorithm, we show the simulation result for the non-orthogonal pilot sequence scenario in Section~\ref{sec5}.} Then, the overall received pilot signal at the $m$-th AP, ${\bY_{m} \in \C^{N \times \tau_p}}$, is obtained as \cite{Marzetta:2010,Ngo:2017}
	\begin{align}
		\bY_{m} = \sum_{k=1}^K \sqrt{\tau_p p_p} \bg_{mk} \bpsi_k^\mathrm{H} +\bW_{m},
	\end{align}
	where
	$p_p$ denotes the pilot transmit power, and ${\bW_{m} \in \C^{N \times \tau_p} }$ denotes the complex additive white Gaussian noise (AWGN) matrix with each element following the independent and identically distribution   $\cC\cN(0, \sigma_n^2)$.
	
	To estimate the channel of $k$-th UE, the $m$-th AP first projects the received pilot signal $\bY_{mk}$ onto the $k$-th UE's pilot sequence to remove the inter-user interference as
	\begin{align}\label{received pilot}
		\by_{mk} =& \bY_m\bpsi_{k} \notag
		\\ =& \sum_{i = 1}^K \left ( \sqrt{\tau_p p_p} \bg_{mi} \bpsi_i^\mathrm{H}\bpsi_k \right ) + \bW_{m}\bpsi_k \notag \\
		=& \sqrt{\tau_p p_p} \bg_{mk} + \bw_{mk},
	\end{align}
	where $\bw_{mk} = \bW_{mk}\bpsi_k$ follows ${\bw_{mk} \sim \cC\cN (0,\sigma_n^2\bI_N)}$ due to $\lVert \bpsi_k \rVert^2 = 1 $.
	Then, the $m$-th AP applies the linear minimum mean square error (LMMSE) estimator, which gives the estimated channel as \cite{linear_estimation}
	\begin{align}\label{input_output2}
		\hat{\bg}_{mk} &= c_{mk} \by_{mk} \notag \\ 
		&= c_{mk}\left(  \sqrt{\tau_p p_p} \bg_{mk} + \bw_{mk} \right),
	\end{align}
	where  $c_{mk} = \sqrt{\tau_p p_p} \beta_{mk}/
	{(\sigma_n^2 + \tau_p p_p\beta_{mk})}$ is the LMMSE estimator. 
	Note that the estimated channel $\hat{\bg}_{mk}$ and the estimation error $\mathbf{e}_{mk}=\bg_{mk} - \hat{\bg}_{mk} $ are independent \cite{Ngo:2017} and satisfy
	\begin{align}
		\hat{\bg}_{mk} &\sim \cC\cN(0,\gamma_{mk}\bI_N ), \\ 
		\quad \mathbf{e}_{mk} &\sim \cC\cN(0,(\beta_{mk}-\gamma_{mk} )\bI_N ), \label{channel error}
	\end{align}
	where $\gamma_{mk} ={\tau_p p_p \beta_{mk}^2}/{(\sigma_n^2+\tau_p p_p \beta_{mk})} $. The estimated channels are used to form a combining filter at the AP during the data transmission phase.
	
	\subsection{Uplink Data Transmission} \label{2-b}
	In this subsection, the uplink data transmission phase is formulated.
	Let $\by_{m}^\mathrm{ul}$ be the uplink received signal at the $m$-th AP, which is represented as
	\begin{align}
		\by_{m}^\mathrm{ul} =  
		\sum_{k=1}^K \sqrt{p_u}\bg_{mk}s_{k} + \bn_m,
	\end{align}
	where $p_u$ denotes the uplink transmit power, $s_{k}$ denotes the UE data symbol that satisfies $\mathbb{E} \{ |s_k|^2\} =1$, and ${\bn_m \sim \cC\cN(0,\sigma_n^2\bI_N)}$ denotes the AWGN.
	
	The maximal-ratio combining (MRC) filter constructed by the estimated channel $\hat{\bg}_{mk}$ is employed at each AP for detecting the data symbol of $k$-th UE. Thus, the processed received signal of the $k$-th UE at the $m$-th AP can be expressed as 
	\begin{align} \label{y_check}
		\check{y}_{mk} = \hat{\bg}_{mk}^\mathrm{H} \by_{m}^\mathrm{ul}=& \hat{\bg}_{mk}^\mathrm{H} \left( 
		\sum_{k'=1}^K \sqrt{p_u}\bg_{mk'}s_{k'} + \bn_m
		\right).
	\end{align}
	
	\textbf{\textit{Remark 1.}}
	In partially centralized cell-free massive MIMO systems, low-level baseband signal processing, e.g., the MRC filter as considered in our work, can be performed directly at the APs \cite{Ngo:2017,Emil:2020}. Since the quantization error occurs in the limited fronthaul link, the pre-processing operations at the APs are not affected by the quantization error.
	

	\subsection{Quantization Model} \label{2-c}

	Because of the limited fronthaul capacity between the APs and CPU, the CPU can only access the quantized version of $\check{y}_{mk}$ while processing all the data. We analyze, therefore, the data quantization at the fronthaul links in this subsection. 
	Since the quantizer input $\check{y}_{mk}$ approximately follows the Gaussian distribution when $K$ is large by the central limit theorem \cite{Bashar:2019:energyefficiency}, we can apply the AQNM on the quantizer \cite{quantization_correlation}, which is regarded as a variant of the Bussgang theorem \cite{bussgang}.
	Then, the AQNM-based quantizer linearizes the nonlinear distortion as  
	\begin{align} \label{quantization}
		\cQ(\check{y}_{mk}) = 
		(1-\rho_{mk})\check{y}_{mk} + n_{mk}^q,
	\end{align}
	where $\rho_{mk}$ denotes the quantization distortion, and $n_{mk}^q$ denotes the quantization noise.
	The quantization distortion $\rho_{mk}$ depends on the allocated bits for the $k$-th UE through the fronthaul link between the CPU and the $m$-th AP. In particular, we use the uniform quantizer whose output has the minimum distortion \cite{Max:1960}, and Table~\ref{quantizerNMSE} gives its $\rho_{mk}$ values under the Gaussian input assumption.
	
		\begin{table}[t]
		\centering
		\captionsetup{justification=centering, labelsep=newline, font={smaller,sc}}
		\caption{Quantization Distortion Value \cite{Max:1960}}
		\label{quantizerNMSE}
		\begin{tabular}{cc}
			\toprule
			$\text{Bits}$  & Quantization distortion $\rho$  \\
			\midrule
			\hline
			\addlinespace
			1  & 0.3634 \\
			2   & 0.1188 \\
			3   & 0.03744  \\
			4  & 0.01154   \\
			5   & 0.003490  \\
			\bottomrule
		\end{tabular}
	\end{table}
	
	From the analyses in \cite{quantization_correlation}, the variance of the quantization noise $n_{mk}^q$ can be approximated as 
	\begin{align} \label{quaantization distortion}
		R_{n_{mk}^q} 
		&= \mathbb {E}\left \{{\left |{n_{mk}^q}\right |^{2}}\right \} \notag
		\\ &\approx \rho_{mk}(1-\rho_{mk})\mathbb{E} \left \{ \check{y}_{mk} \check{y}_{mk}^* \right \} 
		.
	\end{align}
	Note that $n_{mk}^q$ would not necessarily follow the Gaussian distribution, but we assume the Gaussian quantization noise in our scenario to describe the worst case, which provides the lower bound of the system performance\cite{information_theory}. 
	\noindent 
	With the AQNM-based quantizer, the quantization noise and the input signal become uncorrelated, i.e.,
	\begin{align}
		\label{uncorrelated}
		\mathbb{E} \left\{  \check{y}_{mk}^* n_{mk}^q  \right\}
		= 0.
	\end{align}
	Also, the quantization error at one AP does not intrinsically impact the errors at the others, which leads to
	\begin{align}
		\mathbb{E} \left\{  n_{mk}^{q*} n_{m'k}^q  \right\}
		= 0, \text{    if   } m \ne m'.
	\end{align}
	
	\textbf{\textit{Remark 2.}}
In this paper, we consider the scalar qunatizaion instead of vector qunatization since using vector quantization might be infeasible in practical scenarios \cite{scalarquantizer1,scalarquantizer2,scalarquantizer3}. The scalar quantization, however, can be simply designed with slight loss of performance. Furthermore, our proposed algorithm individually allocates fronthual bits for each AP, which cannot be applied in the vector qunatizaion operation.


	\subsection{CPU Receiver Filter} \label{2-d}
	After receiving the quantized received signals $\cQ(\check{y}_{mk})$ from all APs, the CPU linearly combines them to detect the $k$-th UE signal as
	\begin{align}
		\label{filtered}
		r_k&=\sum_{m=1}^M u_{mk} \cQ(\check{y}_{mk}) \notag
		\\ &= \sum_{m=1}^M u_{mk}\left((1-\rho_{mk})
		\hat{\bg}_{mk}^\mathrm{H} \by_{m}^\mathrm{ul} + n_{mk}^q \right ),
	\end{align}
	where  $u_{mk}$ is the receiver filter coefficient for the quantized signal from the $m$-th AP. The filter for the $k$-th UE $\bu_k = [u_{1k},u_{2k}, \cdots, u_{Mk}  ]^\mathrm{T} \in \C^{M \times 1}$ is normalized as $\left \Vert \bu_k \right \Vert^2 =1$ to eliminate the scaling factor ambiguity. We proposed how to design $\bu_k$ in Section~\ref{problem_solving}.

	\section{User Spectral Efficiency} \label{sec3}
	In this section, we obtain the closed-form expression of the SINR 
	to analyze the SE in terms of the fronthaul bit $b_{mk}$ and the receiver filter $\bu_k$.
	Compared to the conventional approach in \cite{Bashar:2019:energyefficiency} that allocates same bits for all APs, our approach allocates distinct bits for each AP and UEs.
	With the restricted acquisition of the channel state information in reality, we assume that the CPU can access the long-term channel statistics only. Then, the filtered $k$-th user signal \eqref{filtered} can be represented as
	
	
	\begin{figure*}[t]		
		\begin{equation}
			\label{SINR}
			\mathrm{SINR}_k =
			\frac{|r_k^D|^2}
			{\mathbb {E} \left \{\left |{ r_k^B }\right | ^{2}\right \}
				+ \sum_{k' \ne k}^K
				\mathbb {E} \left \{\left |{ r_{k'}^I}\right | ^{2}\right \}
				+
				\mathbb {E} \left \{\left |{ r_k^N }\right | ^{2}\right \}
				+
				\mathbb {E} \left \{  |{ r_k^Q } | ^{2}\right \}
			} \tag{17},
		\end{equation}
		\begin{equation}
			\label{SINR1}
			\mathrm{SINR}_k
			=
			\frac{p_u N^2  \bu_k^\mathrm{H}(\bOmega_k^\mathrm{H}\bgamma_k\bgamma_k^\mathrm{H}\bOmega_k)\bu_k}
			{
				\bu_k^\mathrm{H} \Big ( p_u N^2 \bGamma_k^2
				+ 
				\sum_{k' = 1}^K p_uN \bD_{kk'}
				+
				\sigma_n^2N\bGamma_k
				)\bOmega_k\bu_k
				-
				\bu_k^\mathrm{H} \bOmega_k^\mathrm{H} \Big ( p_uN^2\bGamma_k^2
				\Big )\bOmega_k\bu_k
			}.\tag{18}
		\end{equation}
		\hrule
	\end{figure*}
	
	\begin{align}
		\label{decomposed}
		r_k=& \sum_{m=1}^M u_{mk}\left((1-\rho_{mk})
		\hat{\bg}_{mk}^\mathrm{H} \by_{m}^\mathrm{ul} + n_{mk}^q \right ) \notag 
		\\
		=& ~ \underbrace{ \mathbb{E} \left [ \sum_{m=1}^M
			\tilde{u}_{mk}\sqrt{p_u} \hat{\bg}_{mk}^\mathrm{H} {\bg}_{mk}   \right ]}_\mathrm{Desired ~ Signal}s_k  \notag 
		\\
		+& \underbrace{ \sqrt{p_u} \Biggl ( \sum_{m=1}^M
			\tilde{u}_{mk} \hat{\bg}_{mk}^\mathrm{H} {\bg}_{mk} -\mathbb{E} \left [ \sum_{m=1}^M \tilde{u}_{mk} \hat{\bg}_{mk}^\mathrm{H} {\bg}_{mk} \right ] \Biggr )
		}_\mathrm{Beamforming ~ Uncertainty} s_k \notag
		\\
		+& \sum_{k' \ne k}^K \underbrace{\sqrt{p_u} \sum_{m=1}^M  \tilde{u}_{mk}\hat{\bg}_{mk}^\mathrm{H}\bg_{mk'}
		}_\mathrm{Inter-User ~ Interference} s_{k'} \notag
		\\+& \underbrace{\sum_{m=1}^M \tilde{u}_{mk}\hat{\bg}_{mk}^\mathrm{H}\bn_{m}}_\mathrm{Noise} + \underbrace{\sum_{m=1}^M u_{mk}n_{mk}^q}_\mathrm{Quantization ~ Noise},
	\end{align}
	where $\tilde{u}_{mk} = u_{mk} (1-\rho_{mk})$. 
	
	Let $r_{k}^D, r_k^B, r_{kk'}^I, r_k^N$ and $r_k^Q$ be desired signal, beamforming uncertainty, inter-user interference, noise, and quantization noise terms of the $k$-th UE signal in \eqref{decomposed}, respectively. Then, the filtered signal $r_k$ can be equivalently expressed as
	
	\begin{align} \label{receive_abstract}
		r_k = r_k^D s_k + r_k^B s_k +\sum_{k'\ne k}^K r_{kk'}^I s_{k'} + r_k^N + r_k^Q.
	\end{align}
	Using the fact that the channels, noise, and quantization noise in \eqref{receive_abstract} are independent, i.e.,  $r_{k}^D, r_k^B, r_{kk'}^I, r_k^N$, and $r_k^Q$ are mutually uncorrelated, the SINR of the $k$-th user is formulated as \eqref{SINR} shown at the top of this page. Appendix A derives \eqref{SINR1} from \eqref{SINR}.
	The variable expressions in \eqref{SINR1} are denoted as follows,
	\setcounter{equation}{18}
	\begin{align}
		&\bgamma_k =  [\gamma_{1k}, \cdots, \gamma_{Mk}]^\mathrm{T} \in \C^{M \times 1}, \label{variable1} 
		\\ &\bGamma_k = \mathrm{diag} \left([\gamma_{1k}, \cdots, \gamma_{Mk}]^\mathrm{T} \right)\in \C^{M \times M}, \label{variable2} 
		\\ &\bOmega_k \: \, = \mathrm{diag}\left([(1-\rho_{1k}),\cdots, (1-\rho_{Mk})]^\mathrm{T} \right) \in \mathbb{R}^{M \times M}, \label{variable3} 
		\\ &\bD_{kk'} = \mathrm{diag}\left([\gamma_{1k}\beta_{1k'},\cdots, \gamma_{Mk}\beta_{Mk'}]^\mathrm{T} \right) \in \C^{M \times M}. \label{variable4} 
	\end{align}
	Subsequently, the SE of the $k$-th UE under the knowledge of channel statistics is given by 
	\begin{align} \label{SE}
		S_k = \log_2(1+\mathrm{SINR}_k),
	\end{align}
	which is the function of $\bOmega_k$ and $\bu_k$ that are related to the fronthaul bit allocation and the receiver filter design, respectively.

	\section{Receiver Filter Design and Bit Allocation} \label{sec4}
	In this section, we first formulate the total SE maximization problem to investigate how the receiver filter coefficients and the number of allocated fronthaul bits affect the system performance. 
	Then, we propose the HS-based fronthaul bit allocation algorithm to handle the combinatorial non-convex problem.
	The proposed algorithm has a hierarchical structure that first allocates the fronthaul bits for APs and then reallocates the bits for the associated UEs. 
	We refer to the bit allocation algorithm for APs as ``Stage 1" and the bit allocation algorithm for UEs as ``Stage 2."
	Additionally, the analyses on computational complexity and convergence are provided to show the effectiveness of the proposed algorithm.
	We also demonstrate that the proposed algorithm can adapt to other performance metrics, such as the max-min fairness, with minimal adjustments.
	\subsection{Total Spectral Efficiency Maximization}
	We consider the total SE, which is the sum of all the individual SE defined in \eqref{SE}, as our objective in this subsection, while taking into account the constraints on fronthaul bit allocation and the receiver filter design that were previously described.
	Then, the overall problem can be formulated as
	
	\begin{align}
		\mathrm{(P1)}~:~ \max _{\cU,\cB }&~   \sum_{k=1}^K S_k \left ({\mathbf {u}_k,\bb_k }\right)  \notag
		\\\mathrm {s.t. }~&||\mathbf {u}_{k}||=1,~ \forall k,\quad \tag{1a} \label{1-a}
		\\&	\sum_{m=1}^M\sum_{k=1}^K b_{mk}  \le b_\mathrm{max}, \tag{1b} \label{1-c}
		\\&	b_{mk} \in \mathbb{Z}_{0+},~~ \forall m,k, \tag{1c} \label{1-d}
	\end{align}
	where the optimization variables $\cU$ and $\cB$ are defined as the sets of receiver filter coefficients $\cU = \{\bu_1, \cdots, \bu_K \} $ and fronthaul bits $\cB =  \{\bb_1, \cdots, \bb_K\}$, respectively. Here, each vector has the following elements $\bu_k = [u_{1k}, \cdots, u_{Mk}]^\mathrm{T}$ and  $\bb_k = [b_{1k}, \cdots, b_{Mk}] ^\mathrm{T}$.
	The receiver filter coefficients are normalized as in \eqref{1-a} to find the unique solution. Also, the limited capacity of the system is interpreted by the total fronthaul bit condition as \eqref{1-c}, while the bits being forced to have non-negative integer values through \eqref{1-d}, where $\mathbb{Z}_{0+}$ denotes the set of non-negative integers.
	
	Note that the optimization problem $(\mathrm{P1})$ is not convex with respect to $\cU$ and $\cB$. In addition, the optimization variable $\bb_k$ is defined over the discrete domain set, which introduces a combinatorial aspect to $(\mathrm{P1})$. As a result, common convex optimization techniques like gradient descent cannot be used to solve $(\mathrm{P1})$ directly.
	
	\subsection{Proposed Algorithm - Stage1} \label{problem_solving}
	To tackle the combinatorial non-convex problem $(\mathrm{P1})$, we propose a meta-heuristic approach, especially based on the HS algorithm, which provides a near-optimal solution with low computational complexity\cite{Geem:2001}. 
	In this subsection, we describe Stage~1 of the proposed algorithm along with a brief introduction to the core principles of the HS algorithm.
	In Stage~1, we assume that the UEs associated with the same APs have a uniform number of bits, which can initially find a rough bit allocation configuration with low complexity. Thus, we introduce a new notation $\bb_\mathrm{AP} = [b_{\mathrm{AP}1}, \cdots, b_{\mathrm{AP}M}]$ instead of using ${\cB =  \{\bb_1, \cdots, \bb_K\}}$, and the following relationship is held $b_{m1}  = \cdots= b_{mK} =b_{\mathrm{AP}m}$.  
	\noindent
	As with numerous other meta-heuristic algorithms, the HS algorithm consists of two distinct phases: initialization and improvisation.
	
	\textit{1) Initialization:} In the initialization phase, the HS algorithm creates a set of feasible points known as ``harmony memory (HM)." In our proposed algorithm, the HM is defined as
	\begin{align}
		\bPhi=\left[\begin{array}{cccc}
			{ b_{1,\mathrm{AP}1}} & \cdots & b_{1,\mathrm{AP}M} & {E_{1}} \\
			\vdots & \ddots & \vdots & \vdots \\
			{b_{N_\mathrm{HM}^{(1)},\mathrm{AP}1}} & \cdots & {b_{N_\mathrm{HM}^{(1)}, \mathrm{AP}M}} & {E_{N_\mathrm{HM}^{(1)}}}
		\end{array}\right], \label{HM init}
	\end{align}
	where $N_\mathrm{HM}^{(1)}$ is a design parameter for the number of starting points in the algorithm, and the superscript $(\cdot)^{(1)}$ represents that it is the parameter used for Stage~1. Each row of the HM, which is called harmony and denoted by a notation $\bphi_i$, consists of two parts; a variable part and an evaluation part, containing $M$ fronthaul bit values for each AP $\bb_{i,\mathrm{AP}}= [b_{i,\mathrm{AP}1},\cdots,b_{i,\mathrm{AP}M} ]^\mathrm{T}$ and its corresponding performance evaluation value $E_i$, respectively.\footnote{To make the notation consistent, we denote the column vector $\bb_{i,\mathrm{AP}}$ as the $i$-th row of $\bPhi$.}

	During the initialization phase, $\bb_{i,\mathrm{AP}}$ is randomly generated under the total fronthaul bit constraint. Note that, due to the same bit assumption, the relaxed condition can be considered in each harmony, which is given as  
	\begin{equation} 
		\sum_{m=1}^M b_{i,\mathrm{AP}m} \le \frac{b_{max}}{K}. \label{eq: stage1bmax}
	\end{equation}
	Once the variable part of the harmony is fixed as $\bb_{i,\mathrm{AP}}$, we can obtain the evaluation part $E_i$ by substituting $\bb_{i,\mathrm{AP}}$ into the problem $(\mathrm{P1})$ and calculating the total SE.
	Solving $(\mathrm{P1})$, however, requires not only the fronthaul bits  but also the receiver filter coefficients as shown in \eqref{SINR1} and \eqref{SE}. Thus, designing the optimal receiver filter, which can maximize the total SE, is essential. 
	Fortunately, the fixed fronthaul bits mitigate the difficulty of solving the combinatorial nature in $(\mathrm{P1})$, so it can be boiled down to the receiver filter design problem.
	Since the individual SE is the function of its own receiver filter design only,  i.e., $\bu_{k'}$ does not affect $S_k$ for $k' \ne k$, the receiver filters can be designed separately. As a result, the problem  $(\mathrm{P1})$ can be separated into $K$ sub-problems in which each of them uses the individual SE as the objective, and the $k$-th receiver filter design problem given $\bb_i$ can then be represented as
	\begin{align}
		\mathrm{(P1.k)}~:~ \max _{\mathbf {\bu_k} }&~  S_k  =  \log_2 \left ( 1+\mathrm{SINR}_k \left ({\mathbf {u}_k,\bb_{i,\mathrm{AP}} }\right) \right) \notag
		\\\mathrm {s.t. }~& ||\mathbf {u}_{k}||=1,\quad \tag{2a} \label{1-a-k}
	\end{align}
	where the constraint \eqref{1-a-k} denotes the normalization only for the $k$-th receiver filter. 
	In fact, maximizing $\mathrm{SINR}_k$ becomes equivalent to maximizing the $k$-th UE's SE because of the monotonicity of the logarithm in the SE equation.
	Hence, after we rearrange the numerator and denominator of $\mathrm{SINR}_k$ in \eqref{SINR1}, the equivalent problem of $\mathrm{(P1.k)}$ is given by
	\begin{align}
		\mathrm{(P2.k)}~:~ \max _{\mathbf {u}_k }&~   
		\frac{\bu_k^\mathrm{H} \bA_k \bu_k}{\bu_k^\mathrm{H} \bB_k \bu_k}  \notag
		\\\mathrm {s.t. }~& \eqref{1-a-k}, \notag
	\end{align}
	with the following auxiliary variables
	\begin{align}
		\mathbf {A}_{k} &= N\bOmega_i^{ \mathrm{H}}\bgamma_k\bgamma_k^\mathrm{H}\bOmega_i, \label{auxA} \\
		\mathbf {B}_{k} &= {	\Big ( N(\bI_M - \bOmega_i^{\mathrm{H}})\bGamma_k^2 + \sum_{k' = 1}^K  \bGamma_{k}\bD_{kk'}	+\frac{\sigma_n^2}{p_u}\bGamma_k	\Big )\bOmega_i},\label{auxB}
	\end{align}
	where $\bOmega_i$ is obtained from \eqref{variable3} by computing the quantization distortion values corresponding to $\bb_{i,\mathrm{AP}}$ and independent with the UE index due to same bits are allocated for UEs.

	The matrices $\bGamma_k$, $\bOmega_i$, and $\bD_{kk'}$ are Hermitian because they are diagonal matrices, as defined in \eqref{variable2}, \eqref{variable3}, and \eqref{variable4}, respectively. Since all matrices comprising $\mathbf {A}_{k}$ and $\mathbf {B}_{k}$ are Hermitian, and the elements in $\bOmega_i$ range from $0$ to $1$ regardless of $\bb_i$ selection, it follows that $\mathbf {A}_{k}$ is Hermitian and $\mathbf {B}_{k}$ is positive definite. As a result, the SINR maximization problem $(\mathrm{P2.k})$ becomes the generalized eigenvalue problem, which is also called as the generalized Rayleigh quotient \cite{Bashar:2019:energyefficiency,Gene:2013}. Then, the optimal receiver filter is obtained as 
	\begin{align} \label{eq: u_k}
		\bu_k  = \eta \operatorname*{argmax}_{\bv} \{\lambda : \bT\bv = \lambda\bv, \bT = \left ( \bB_k\right)^{-1} \bA_k\},
	\end{align}
	where $\eta$ is the normalization factor to satisfy the constraint \eqref{1-a-k}.
	With the optimal receiver filter, we can obtain the $i\text{-th}$ evaluation part by computing ${E_i = \sum_{k=1}^K S_k \left ({\bu_{ik},\bb_{i,\mathrm{AP}} }\right)}$.
	To avoid confusion, we use the notation $\cU_i  = \{\bu_{i1}, \cdots, \bu_{iK}\}$ instead of $\cU$, which ensures that the receiver filters are optimally designed only for given $\bb_{i,\mathrm{AP}}$.
	
	
	
	
	\textit{2) Improvisation:} So far, we have discussed how to calculate the optimal receiver filter and its evaluation value given the allocated fronthaul bits. To achieve better performance, however, it is needed to update the fronthaul bit allocation through the improvisation phase of the HS algorithm. In the improvisation phase, we generate new harmonies that can potentially outperform the existing harmonies. The proposed algorithm constructs a new harmony by using a design parameter, denoted as \( D^{(1)} \), which we term the harmony memory considering rate (HMCR). This parameter lies between 0 and 1 and acts as a threshold to decide how the new harmony is formed.
	A newly generated harmony can be represented as	
		\begin{equation}
		\bb_\mathrm{new,AP} = 
		\begin{cases} 
			[b_{\mathrm{new,AP}1} , \cdots , {b_{\mathrm{new,AP}M}} ] & \text{w/ prob. } 1-D^{(1)}, \\
			[b_{\alpha_1,\mathrm{AP}1} , \cdots , {b_{\alpha_M, \mathrm{AP}M}} ] & \text{w/ prob. } D^{(1)}.
		\end{cases} \label{eq: satge1newharmony}
	\end{equation}
	When the randomly chosen probability is greater than $D^{(1)}$, we randomly generate a new harmony similar to the initialization phase. If it is less than or equal to $D^{(1)}$, we build the new harmony using bits  already exist in the HM. Each $\alpha_m$ represents a harmony index, chosen randomly from the set $\{1,2, \cdots, N_\mathrm{HM}^{(1)}\}$.  Thus, the new harmony is more likely to include the fronthaul bits that occur frequently in the HM. Note that  the total bit constraint should be satisfied in constructing new harmonies. If the new harmony exceeds this limit, we iteratively subtract one bit from a randomly chosen position with non-zero bits until the constraint is met.
	
	The newly obtained evaluation value $E_\mathrm{new}$ of the new harmony can also be computed by solving $(\mathrm{P2.k})$ with the new fronthaul bits $\bb_\mathrm{new,AP}$.
	After calculating $E_\mathrm{new}$, we update the HM by comparing the existing harmonies with the new one. Without loss of generality, we assume that the HM is arranged in decreasing order with respect to the evaluation values as
	\begin{align}
		E_1 \ge E_2 \ge \cdots \ge E_{N_\mathrm{HM}^{(1)}}.
	\end{align}
	Then, the ${N_\mathrm{HM}^{(1)}}$-th harmony is considered as the worst harmony, and we compare $E_\mathrm{new}$ and $E_{N_\mathrm{HM}^{(1)}}$  to select better harmony.
	If $E_\mathrm{new} > E_{N_\mathrm{HM}^{(1)}}$, the updated HM $\bPhi'$ will contain the new harmony after discarding the ${N_\mathrm{HM}^{(1)}}$-th harmony, which is given by
	\begin{align}
		\bPhi'=\left[\begin{array}{cccc}
		{ b_{1,\mathrm{AP}1}} & \cdots & b_{1,\mathrm{AP}M} & {E_{1}} \\
		\vdots & \ddots & \vdots & \vdots \\
		{b_{N_\mathrm{HM}^{(1)}-1,\mathrm{AP}1}} & \cdots & {b_{N_\mathrm{HM}^{(1)}-1, \mathrm{AP}M}} & {E_{N_\mathrm{HM}^{(1)}-1}} \\
			{b_{\mathrm{new,AP}1}} & \cdots & {b_{\mathrm{new,AP}M}} & {E_{\mathrm{new}}}
		\end{array}\right]. \label{HMupdate}
	\end{align}
	Otherwise, if $E_\mathrm{new} \le E_{N_\mathrm{HM}^{(1)}}$, the existing HM will be maintained without replacement, i.e., $\bPhi' = \bPhi$.
	In the proposed algorithm, the process of improvisation continues until the predefined number of iterations $N_\mathrm{iter}^{(1)}$ is reached. By sorting the HM in descending order, we can find the best fronthaul bit allocation strategy and its total SE by selecting the first harmony that consists of $\bb_{1,\mathrm{AP}}$ and its corresponding evaluation $E_1$ from the lastly updated HM.
	The overall algorithm of Stage~1 is summarized in Algorithm~\ref{HS algorithm}.

\begin{algorithm}[t]
				
	\caption{Proposed Algorithm - Stage~1}
	\label{HS algorithm}
	\begin{algorithmic}[1]
		\State \textbf{Initialization}: for a given $b_\mathrm{max}$

		\For {$i = 1,2 \cdots, N_\mathrm{HM}^{(1)}$}
		
		\State Generate $\bb_{i,\mathrm{AP}}$ satisfying constraints \eqref{eq: stage1bmax}
		
		\State Calculate $\bu_{ik} \; \forall k=1,2,\cdots, K$ using \eqref{eq: u_k}
		
		\State Evaluate the $i$-th harmony using \eqref{SE}

		\EndFor
		
		\State Sort $\bPhi$ in descending order with evaluation values
		
		\State \textbf{Improvisation}: for a given $D^{(1)}$
		
		\For {$\mathrm{iter}=1,2,\cdots, N_\mathrm{iter}^{(1)}$}  
		
		\State Generate $\bb_{\mathrm{new,AP}}$ using \eqref{eq: satge1newharmony}
		\State Calculate $\bu_{\mathrm{new},k} \; \forall k=1,2,\cdots, K$ using \eqref{eq: u_k}
		\State Evaluate the new harmony using \eqref{SE}
		
		\State \textbf{If} $E_\mathrm{new} > E_{N_\mathrm{HM}^{(1)}}$
		\State \quad Replace $\bphi_{N_\mathrm{HM}^{(1)}}$ with $\bphi_{\mathrm{new}}$

		\State \textbf{End if}
		\State Sort $\bPhi'$ in descending order with evaluation values 
		\EndFor
		
		\State \textbf{Output:}
		\State Select $E_1$ to evaluate the performance of Stage~1
		\State Choose $\bb_{1,\mathrm{AP}m}$ as the best AP bit allocation configuration
		
	\end{algorithmic}
	
\end{algorithm} 

	%
	
			\begin{algorithm}[t]
		\caption{Proposed Algorithm - Stage~2}
		\label{HS algorithm2}
		\begin{algorithmic}[1]
			\State \textbf{Outer Loop}: for a given $N_\mathrm{iter}^o$
			
			\While{outer loop iteration number not met $N_\mathrm{iter}^o$}
			
			\For{$m = 1,2 \cdots, M$}
			
			
			\State \textbf{Initialization}: for a given $b_{1,\mathrm{AP}m}$

			\For {$j = 1,2 \cdots, N_\mathrm{HM}^{(2)}$}
			
			\State Generate $\bb_{j,m}$ satisfying constraints \eqref{eq: stage2bmax}
			
			\State Calculate $\bu_{jk} \; \forall k=1,2,\cdots, K$ using \eqref{eq: u_k}
			
			\State Evaluate the $j$-th harmony using \eqref{SE}

			\EndFor
			
			\State Sort $\bPi_m$ in descending order with evaluation
			
			\; \; values 
			
			\State \textbf{Improvisation}: for a given $D^{(2)}$
			
			\For {$\mathrm{iter}=1,2,\cdots, N_\mathrm{iter}^{(2)}$}  
			
			\State Generate $\bb_{\mathrm{new},m}$ using \eqref{eq: stage2newharmony}
			\State Calculate $\bu_{\mathrm{new},k} \; \forall k=1,2,\cdots, K$ using \eqref{eq: u_k}
			\State Evaluate the new harmony using \eqref{SE}
			
			\State \textbf{If} $E_{\mathrm{new},m} > E_{N_\mathrm{HM},m}$
			\State \quad Replace $\bpi_{N_\mathrm{HM}^{(2)}}$ with $\bpi_{\mathrm{new}}$

			\State \textbf{End if}
			\State Sort $\bPi_m'$ in descending order with evaluation 
			
			\quad \quad \; values 
			\EndFor

			\EndFor
			
			\EndWhile
			\State \textbf{Output:}
			\State Select $E_{1,M}$ to evaluate the performance of Stage~2
			\State Choose the best bit allocation configuration
			
		\end{algorithmic}
		
	\end{algorithm} 
	
	\subsection{Proposed Algorithm - Stage2} \label{prop stage2}

		In Stage~2, the bit allocation configuration is fine-tuned by reallocating different bits to the UEs. Since the same bit allocation might be inefficient for specific UEs who have unfavoralbe channel qualities, the algorithm in Stage~2 effectively adjusts their bits to enhance its performance. Our main approach involves reallocating bits among UEs within a particular AP, iterating through each AP in turn. Since all the UE-AP pairs are taken into account for bit allocation, we utilize the notation ${\cB =  \{\bb_1, \cdots, \bb_K\}}$ in Stage~2. With this extended consideration, we again construct the harmonies and update the HM by comparing their evaluation values.
		In Stage~2, the total bit constraint considered at the $m$-th AP is represented as  
		\begin{equation}
			\sum_{k=1}^K b_{mk} \le Kb_{1,\mathrm{AP}m}, \label{eq: stage2bmax}
		\end{equation}
		where $b_{1,\mathrm{AP}m}$ denotes the best allocated bits for the $m$-th AP harmony resulted from Stage~1, and $K$ times of this implies entire available bits in the AP. 
		
		When we consider bit allocation within the $m$-th AP in Stage~2, the HM is initially constructed as follows 
		\begin{align}
			\bPi_m=\left[\begin{array}{cccc}
				b_{1,\{m,1\}}& \cdots& b_{1,\{m,K\}}& E_{1,m} \\
				\vdots & \ddots & \vdots & \vdots \\
				b_{N_\mathrm{HM}^{(2)},\{m,1\}}& \cdots& b_{N_\mathrm{HM}^{(2)},\{m,K\}}& E_{N_\mathrm{HM}^{(2)},m}
			\end{array}\right], \label{HM init2}
		\end{align}
		where the harmonies have $K$ elements corresponding to the number of UEs.
		Same as Stage~1, the $j$-th harmony consists of variable part and evaluation part $\bpi_{j,m} = [b_{j,\{m,1\}}, \cdots, b_{j,\{m,K\}}, E_{j,m}]$, where the subscripts $j$ and $m$ are the harmony index and AP index, respectively. 
		In the improvisation phase, a variable part of new harmony is generated as
		\begin{align} \label{eq: stage2newharmony}
			&\bb_{\mathrm{new},m}  \notag  \\& = 
			\begin{cases} 
				[b_{\mathrm{new},\{m,1\}} , \cdots, {b_{\mathrm{new},\{m,K\}}} ] & \text{w/ prob. } 1-D^{(2)}, \\
				[b_{\tilde{\alpha}_1,\{m,1\}} ,\cdots, {b_{\tilde{\alpha}_K, \{m,K\}}} ] & \text{w/ prob. } D^{(2)},
			\end{cases} 
		\end{align}
		where each $\tilde{\alpha}_{k}$ represents a harmony index.
		Note that Stage~2 only focuses on a specific AP, which treats the remaining bits in other APs as constants. 
		Then, the algorithm updates the HM with better harmonies based on the evaluation values that are computed by \eqref{eq: u_k} with their allocated bits.  
		When bit allocation is complete on one AP, the algorithm moves to the next AP and repeats the above procedure. Since the bit allocation on the other AP disturbs the bit allocation result of the previous AP, to reflect the effect of bit configuration changes from other APs, we introduce an outer cycle of Stage~2, denoted as $N_\mathrm{iter}^o$. The output of Stage~2 shows the bit allocation configuration between all the UE-AP connections. The entire process of Stage~2 is summarized in Algorithm~\ref{HS algorithm2}.

	\subsection{Convergence and Complexity Analyses} \label{convergence}
	In this subsection, we analyze the convergence and computational complexity of our proposed algorithm. Since the entire procedures of Stage~1 and Stage~2 are identical except for the numbers of iterations, we only show the convergence of Algorithm~\ref{HS algorithm} to avoid redundancy. We first define 
	$f(\bb_\mathrm{AP} ; \bPhi)$ as the objective function of the problem $\mathrm{(P1)}$, where $\bb_\mathrm{AP}$ denotes the set of allocated fronthaul bits in the HM $\bPhi$. Note that, the receiver filter design also affects the objective but can be always found when the fronthaul bits are fixed, so we only focus on the variable $\bb_\mathrm{AP}$ for a simpler analysis.
	At the $\ell$-th iteration, $\bb_\mathrm{AP}^{(\ell)}$ is optimized by selecting the harmony that has the maximum evaluation value among the $\ell$ times updated HM, which is denoted by $\bPhi^{(\ell)}$. Then, it follows that 
	\begin{align}
		f(\bb_\mathrm{AP}^{(\ell+1)} ; \bPhi^{(\ell+1)}) &= \max_{\{i \vert \bb_{i,\mathrm{AP}} \in \bPhi^{(\ell+1)}\}} \sum_{k=1}^K S_k \left ({\bu_{ik},\bb_{i,\mathrm{AP}}}\right)
		\\
		&\stackrel{(a)}{\ge} \max_{\{i \vert \bb_{i,\mathrm{AP}} \in \bPhi^{(\ell)}\}} \sum_{k=1}^K S_k \left ({\bu_{ik},\bb_{i,\mathrm{AP}} }\right)
		\\&=f(\bb^{(\ell)} ; \bPhi^{(\ell)}),
	\end{align}
	where the set ${\{i \vert \bb_{i,\mathrm{AP}} \in \bPhi^{(\ell)}\}}$ denotes the feasible candidates for the harmony selection at the $\ell$-th iteration,  and $(a)$ holds since the update procedure of the proposed algorithm ensures that the updated HM contains better or equal harmonies than the previous one with respect to the evaluation value. As a result, we can justify that the objective is non-decreasing over the iterations. Since the number of feasible fronthaul bit allocation candidates is finite, we can find the optimal fronthaul bits that provides the upper bound of the performance. Considering non-decreasing property of the objective and the existence of the upper bound, the proposed algorithm can be guaranteed to converge.

	\textbf{\textit{Remark 3.}}
	In the proposed algorithm, the convergence of the function value is closely linked with the progress of the iterations, particularly in the results of fronthaul bit allocation. As the algorithm moves forward, it uses existing harmonies to create better versions and selectively updates the HM with top-performing solutions. This not only pushes the function value towards its peak but also steadily enhances and stabilizes the bit allocation outcomes, showing a reliable pattern of iterative refinement and convergence.

	Next, we analyze the computational complexity of the proposed algorithm. In the proposed algorithm, we assume that the complexity of generating harmonies, e.g., random generation or random selection, can be negligible since most of the computing resources are required for designing the receiver filters. As previously discussed, we can design the optimal receiver filter by solving the generalized eigenvalue problem, which takes  $\cO(M^3)$ computational complexity \cite{Choi:2020}.
	Hence, the computational complexity of Algorithm~\ref{HS algorithm} can be expressed as $\cO\left((N_\mathrm{HM}^{(1)}+N_{\mathrm{iter}}^{(1)})KM^3\right)$, taking into account $K$ independent receiver filter design problems that are solved for each harmony and $(N_\mathrm{HM}^{(1)}+N_{\mathrm{iter}}^{(1)})$ total harmonies that are generated throughout the both phases. Although the harmony of Algorithm~\ref{HS algorithm2} consists of $K$ elements that are corresponding to bits for each UE, the complexity depends on the matrix size that is used to the generalized eigenvalue problem, which results in $M^3$ same as Algorithm~\ref{HS algorithm}. Then, the overall computational complexity of Stage~2 can be computed as $\cO\left(N_\mathrm{iter}^o \times M\times (N_\mathrm{HM}^{(2)}+N_{\mathrm{iter}}^{(2)})KM^3\right)$, considering the numbers of outer loops and APs.
	Notably, the complexity of the proposed algorithm is significantly lower than the complexity of the exhaustive search method that requires $\cO\left(\frac{(KM+b_\mathrm{max}-1)!}{(KM-1)!(b_\mathrm{max})!} KM^3\right)$, where $\frac{(KM+b_\mathrm{max}-1)!}{(KM-1)!(b_\mathrm{max})!}$ represents the total number of possible cases that satisfy the total fronthaul bit constraint in \eqref{bit constaint}. When we focuses on the limited exhaustive search case allocating AP bits only, the number of possible sets is reduced as  $\frac{(M+b_\mathrm{max}/K-1)!}{(M-1)!(b_\mathrm{max}/K)!}$, which leads to the complexity $\cO\left(\frac{(M+b_\mathrm{max}/K-1)!}{(M-1)!(b_\mathrm{max}/K)!} KM^3\right)$.

	

	\subsection{Max-min Fairness} \label{adaptability}
	In this subsection, we investigate how our proposed algorithm can be extended to the max-min fairness.
	Similar to the total SE maximization problem $(\mathrm{P1})$, we can formulate the max-min fairness problem, which is given by 
\begin{align}
	\mathrm{(P3)}~:~ \max _{\cU,\cB }&~   \min_k  S_k \left ({\mathbf {u}_k,\bb_k }\right)  \notag
	\\\mathrm {s.t. }~&||\mathbf {u}_{k}||=1,~ \forall k,\quad \tag{3a} \label{3-a}
	\\&	\sum_{m=1}^M\sum_{k=1}^K b_{mk}  \le b_\mathrm{max}, \tag{3b} \label{3-c}
	\\&	b_{mk} \in \mathbb{Z}_{0+},~~ \forall m,k, \tag{3c}  \label{3-d}
\end{align}
	Since the objective has been changed to maximize the minimum SE, a new algorithm is required to find the optimal fronthaul bits and receiver filters with respect to the new objective. The independence of designing $\bu_k$, however, can be preserved given fronthaul bits, which allows the optimal receiver filters to be obtained by solving the generalized eigenvalue problem as before. Therefore, we can concentrate on how to update the fronthaul bits in the new algorithm. From the fact that the HS-based algorithm uses the objective as the evaluation, we can take the minimum SE as the new evaluation metric, which is represented by 
	\begin{align}
		\begin{cases}
			E_i = \min_k S_k ({\mathbf {u}_{ik} ,\bb_{i,\mathrm{AP}}} ) &\text{ for Stage 1},\\
			E_{j,m}= \min_k S_k ({\mathbf {u}_{jk} ,\{b_{j,mk}\}_{m=1}^M }) &\text{ for Stage 2}.
		\end{cases}
	\end{align}
	The initialization and improvisation phases of the new algorithm are identical to previous ones except for the evaluation, and we only need to take into account the minimum SE during the update process. Similarly, the proposed algorithm can adapt to various objectives by changing the evaluation into the new objective, if the independency of designing the receiver filters is preserved.
	
	
	
\begin{table}[t]
	\centering
	\captionsetup{justification=centering, labelsep=newline, font={smaller,sc}}
	\caption{Simulation Parameters}
	\label{tab:simulation-parameters}
	\small 
	\begin{tabular}{ll}
		\toprule
		Parameter & Value \\
		\midrule
		Carrier Frequency ($f_c$) & $2.1$ GHz \\
		Bandwidth ($\mathrm{BW}$) & $20$ MHz \\
		UE Transmit Power ($p_p$ and $p_u$) & $15$ dBm \\
		Noise Figure ($N_f$) & $9$ dB \\
		Noise Temperature ($T_0$) & $290$ K \\
		Maximum Bits ($b_\mathrm{max}$) & $64$ \\
		Number of APs ($M$) & $4$ \\
		Number of UEs ($K$) & $8$ \\
		Number of Antennas per AP ($N$) & $64$ \\
		& \\
		Harmony Search Parameters - Stage 1 & $N_\mathrm{HM}^{(1)} = 10$\\
		&  $D^{(1)} = 0.9$ \\
		&  $N_\mathrm{iter}^{(1)} = 30$ \\
		& \\
		Harmony Search Parameters - Stage 2 & $N_\mathrm{HM}^{(2)} = 5$\\
		&  $D^{(2)} = 0.9$ \\
		&  $N_\mathrm{iter}^{(2)} = 10$ \\
		&  $N_\mathrm{iter}^o = 2$ \\
		\bottomrule
	\end{tabular}
\end{table}

	\section{Simulation Results} \label{sec5}
\begin{figure}[t]
		\centering
		\includegraphics[width=9 cm]{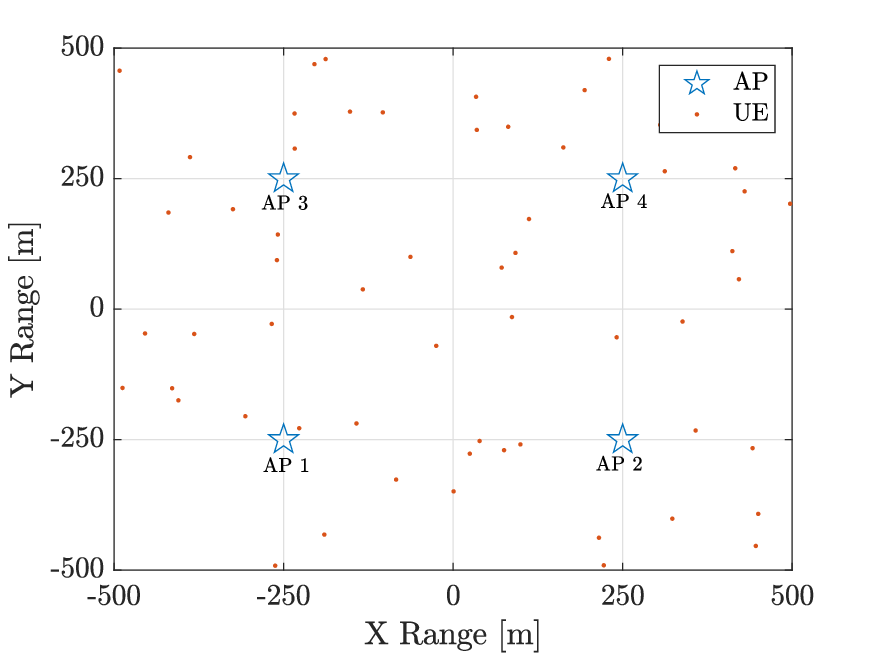}%
		\caption{The APs' locations and the sample of UE scatters in the simulation setup.}
		\label{setup}
	\end{figure}

	In this section, we present simulation results to evaluate the performance of the proposed algorithm. The simulation parameters are set based on the Urban Micro (UMi) scenario in ITU-R M.2412-0 \cite{itu2017guidelines}. The carrier frequency, bandwidth, UE transmit power, and noise figure are set to $f_c = 2.1$ GHz, $\mathrm{BW} = 20$ $\mathrm{MHz}$, $p_p = p_u = 15$ $\mathrm{dBm}$, and $N_f = 9$ $\mathrm{dB}$, respectively. The noise power is set to $\sigma_n^2 = \mathrm{BW} \times k_B \times T_0 \times N_f$, where $k_B = 1.381 \times 10^{-23}$ $\mathrm{J/K}$ represents the Boltzmann constant and $T_0= 290$ $\mathrm{K}$ denotes the noise temperature. Furthermore, we ensure orthogonal pilot sequence transmission by setting $\tau_p = K$. For channel generation, we use the pathloss model (in $\mathrm{dB}$ scale) described in ITU-R M.2412-0 as 
	\begin{align}
		\mathrm{PL} = 36.7\log_{10}(d) + 22.7  + 26\log_{10}(f_c) + \delta_\mathrm{shad},
	\end{align}
	where $d$ is the UE-AP distance in meters and $\delta_\mathrm{shad}$ is the shadow fading factor distributed as $\cN(0,\sigma_\mathrm{shad}^2)$ with the derivative $\sigma_\mathrm{shad} = 4$ $\mathrm{dB}$.
	Unless otherwise specified, the simulation setups are fixed as $b_\mathrm{max}=64$, $M = 4$, $K=8$, and $N=64$, and the design parameters for the proposed algorithm are $N_\mathrm{HM}^{(1)} = 10$, $D^{(1)} = 0.9, N_\mathrm{iter}^{(1)} = 30$ in Stage~1 $N_\mathrm{HM}^{(2)} = 5$, $D^{(2)} = 0.9, N_\mathrm{iter}^{(2)} = 10 $ in Stage~2 , and $N_\mathrm{iter}^o = 2$. The simulation parameters are summarized in Table~\ref{tab:simulation-parameters}. The four APs are positioned at $(\pm 250$~$\mathrm{m},$ $\pm 250$~$\mathrm{m})$, while the UEs are randomly placed within a $1 \times 1$ $\mathrm{km}^2$ square area, as shown in~Fig.~\ref{setup}.

	To compare the proposed algorithm, we evaluate two other benchmarks: AP exhaustive search and equal bit allocation. The AP exhaustive search method examines all possible AP bit allocation sets under the constraint \eqref{eq: stage1bmax}, providing an upper bound on the performance of Algorithm~\ref{HS algorithm}. Only the AP exhaustive search method is considered since computing possible bit allocation configurations for all the UE-AP connections is infeasible. Thus, we show the near-optimal performance of the proposed algorithm by comparing Stage~1 of proposed algorithm and AP exhaustive search. Meanwhile, the equal bit allocation method distributes the limited fronthaul bits equally to all APs and is considered as a lower bound of the performance, which is suitable when the CPU lacks the knowledge of the channel information. 
	
	Our proposed algorithm has three options; \textit{Stage~1}, \textit{Stage~2}, and \textit{Stage~1+2}.  \textit{Stage~1} allocates only AP bits and assumes that UEs have the same amount of bits, whereas \textit{Stage~2} considers a scenario where all the APs have the same capacity that is related to the same total bit constraint. In \textit{Stage~2}, therefore, we do not take into account the AP side bit allocation, and allocates bits for UEs to satisfy \eqref{eq: stage2bmax}, where $b_{1,\mathrm{AP}m}$ has the same value for all $m$ as $b_{1,\mathrm{AP}1}= \cdots = b_{1,\mathrm{AP}M} = \frac{b_{max}}{MK}$. \textit{Stage~1+2} simultaneously utilizes Algorithm~\ref{HS algorithm} and Algorithm~\ref{HS algorithm2}, which first allocates fronthaul bits in AP side and then reallocates them to each UEs. By investigating the performances of three options, we can provide the potential that the proposed algorithm can apply to various scenarios having different resource limitation.
	
	
	\begin{figure}[t]
		\centering
		\includegraphics[width=9 cm]{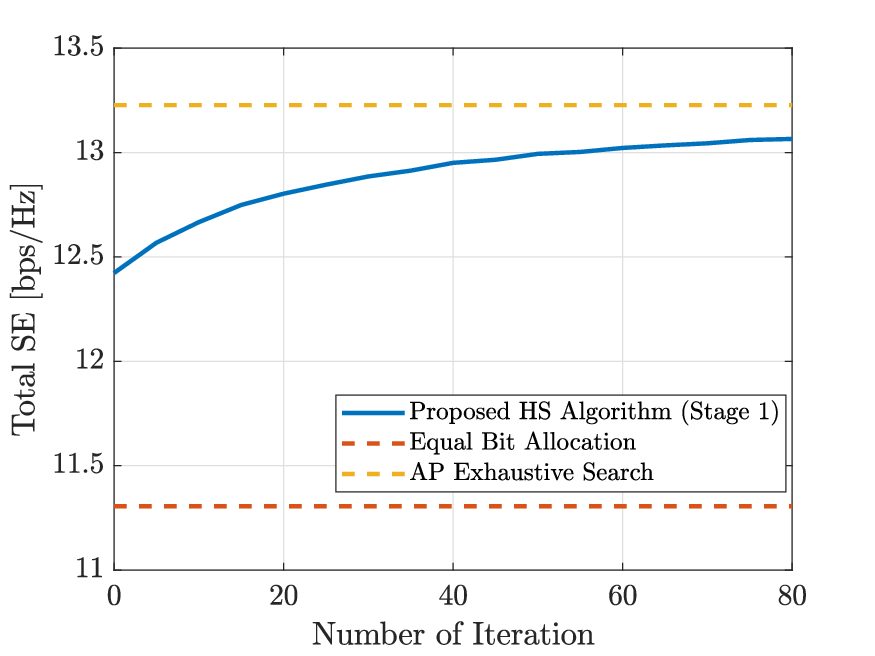}%
		\caption{The empirical convergence of the proposed HS algorithm.}
		\label{empirical convergence}
	\end{figure}

In Fig.~\ref{empirical convergence}, we firstly plot the total SE of the proposed algorithm with the iteration number to demonstrate the convergence of proposed algorithm. Note that only AP exhaustive search scheme can be simulated due to its low complexity. Therefore, we compare Stage~1 of the proposed algorithm and AP exhaustive search. It is shown that the proposed algorithm steadily approaches the optimum performance, which can be obtained via the AP exhaustive search method, over iterations. With this result, we can verify that the proposed algorithm provides empirical convergence, which is consistent with the convergence analysis in Section~\ref{convergence}.

	\begin{figure}[t]
		\centering
		\includegraphics[width=9 cm]{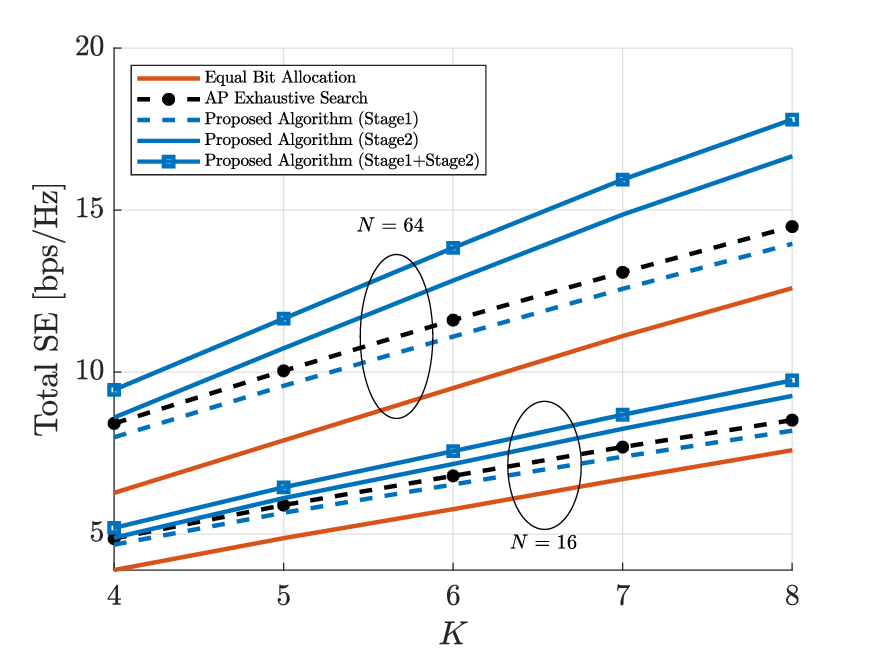}%
		\caption{The total SE versus the number of UEs with the different antenna sizes ($N=$ 16, 64).}
		\label{figure1}
	\end{figure}

	In Fig.~\ref{figure1}, we analyze how the number of UEs affects the total SE. We set the number of UEs to be changed from $K = 4$ to $K = 8$. As shown in the figure, the total SE increases in all the methods as the number of UEs increases, and Stage 1 of the proposed algorithm shows near-optimal performance compared to the AP exhaustive search method. The performance gap between Stage~1 and Stage~2 rises when more UEs are supported. This increasing gap comes form the fact that there is no dominant AP that receives relatively higher signal power than others. Thus, when the number of UEs increases, allocating fronthaul bits within the AP is more effective than AP side bit allocation. It is also verified that the performance of proposed algorithm can be further improved by using both stages.
	We also investigate the performance with respect to the number of antennas that varies from $N=16$ to $N=64$, and it is verified that an increase in $N$ leads to performance enhancement because using more antennas allows the channel between the AP and the UE to have a larger dimension; therefore it facilitates the establishment of favorable channel condition, where the channels are less correlated. 
	As a result, with large $N$, the MRC filter can effectively reduce the inter-user interference and enhance the total SE as well as the SINR of each UE.
	It is observed that when $N$ is large and $K$ is small, proper fronthaul bit allocation leads to more performance improvement than the equal bit allocation.

	\begin{figure}[t]
		\centering
		\includegraphics[width=9 cm]{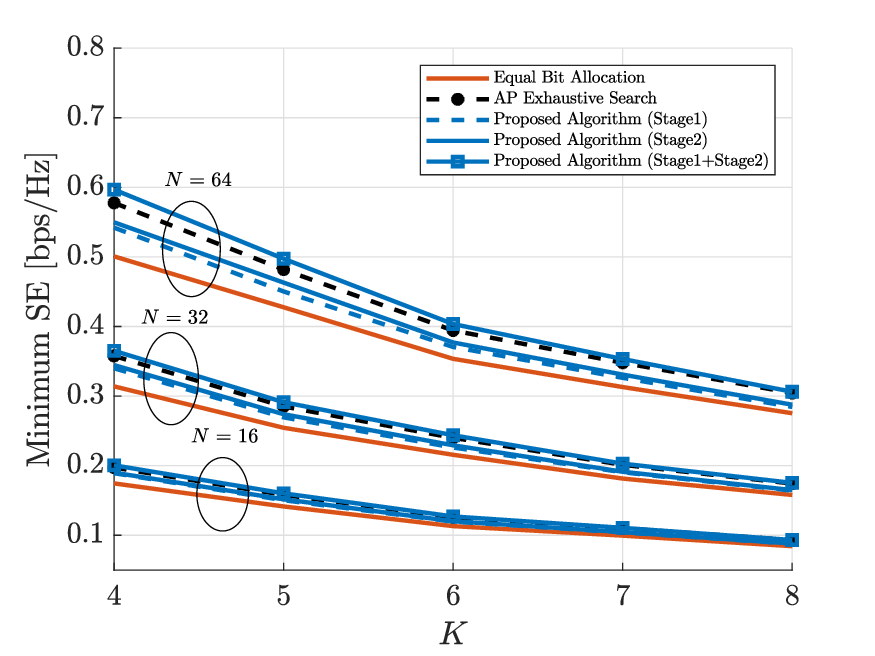}%
		\caption{The minimum SE versus the number of UEs with the different antenna sizes ($N=$ 16, 32, 64).}
		\label{maxmin_user}
	\end{figure}
	
	In Fig.~\ref{maxmin_user}, we evaluate the performance of the proposed algorithm with the max-min fairness metric with respect to the numbers of UEs and antennas as in Fig.~\ref{figure1}. 
	Opposite to the previous figure, the individual SE suffers from the reinforced inter-user interference as the number of UEs increases, resulting in the minimum SE decreasing.
	The minimum SE increases, however, as the number of antennas increases since the expanded antenna dimension leads to the favorable channel condition regardless of the performance metric, providing more fronthaul bit allocation gain. As expected, the proposed algorithms can significantly outperform the lower bound, the equal bit allocation.  The AP exhaustive search scheme, however, provides better performance than Stage~2, and the gap between them increases when APs are equipped with more antennas. This result implies that the inter-user inference at the AP level is more crucial than the inter-user interference at the CPU level for the fairness, because the AP exhaustive search assumes the same bit allocation within the specific AP and Stage~2 assumes different bit allocation for UEs. Still, using both proposed algorithms together results in the best performance among the benchmarks. 
	
	
	
		\begin{table*}[t]
		\renewcommand{\arraystretch}{1.1}
		\centering
		\captionsetup{justification=centering, labelsep=newline, font={smaller, sc}}
		\caption{Comparison of Parameters in Meta-heuristic Algorithms}
		\label{size}
		\begin{tabular}{|l|c|c|c|c|c|c|}
			\hline
			\multirow{2}{*}{Algorithm} & \multicolumn{3}{c|}{Stage 1} & \multicolumn{3}{c|}{Stage 2} \\ \cline{2-7}
			& Population Size  &  Generated Solutions & Iterations & Population Size  &  Generated Solutions & Iterations \\ \hline
			Proposed HS algorithm  & 10  & 1  & 30 &  5 & 1  & 80 \\ \hline
			GA    & 10  &  1  & 30 & 5 & 1  & 80  \\ \hline
			GA (Elitist Selection)  & 10  & 45 & 30  & 5  &  10 &  80 \\ \hline
			Integer PSO    & 10  & 10    & 3   & 5  & 5   & 16 \\ \hline
			Integer PSO (10 fold complexity)   & 10  & 10    & 30   & 5  & 5   & 160 \\ \hline
			SA    &  1  & 1  & 40  &  1  & 1  & 100 \\ \hline
			
		\end{tabular}
	\end{table*}
	
	\begin{figure}[t]
		\centering
		\includegraphics[width=9 cm]{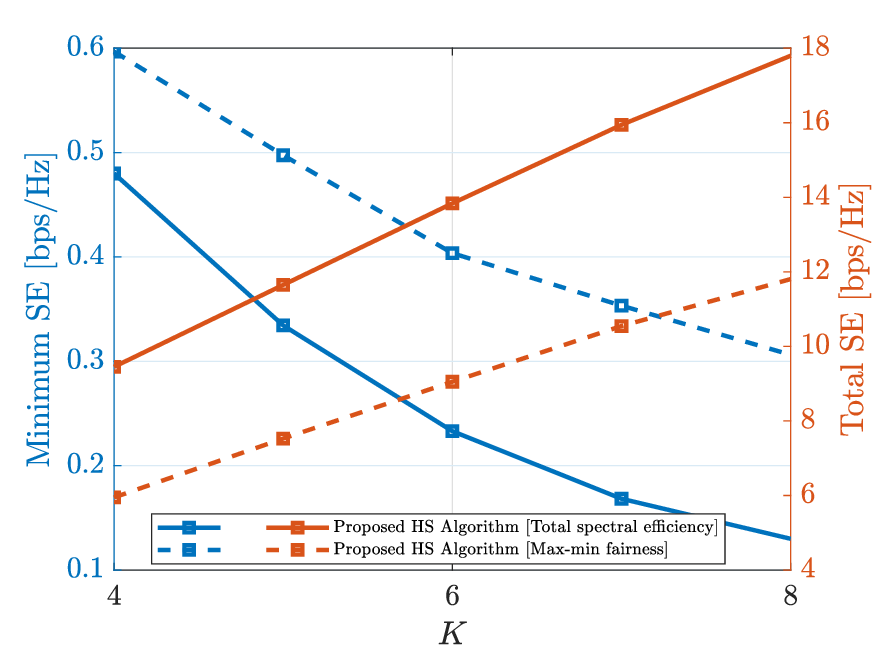}%
		\caption{The total SE and the minimum SE versus the number of UEs with two proposed algorithms that respectively adapt to the performance metrics.}
		\label{compare_maxmin}
	\end{figure}

	In Fig.~\ref{compare_maxmin}, we plot the two figures showing the performance of proposed algorithm with respect to the total SE and the minimum SE.
	According to Fig.~\ref{compare_maxmin}, the results show that the proposed algorithm properly adapts to the considered performance metric. The proposed algorithm for maximizing the total SE tends to allocate more fronthaul bits to the UEs with favorable channel conditions and sacrifice the inferior UEs with small channel gains, whereas the algorithm taking the max-min fairness into account tries to improve the performance of the worst UE to enhance the fairness.

\begin{figure}[t]
	\centering
	\includegraphics[width=9 cm]{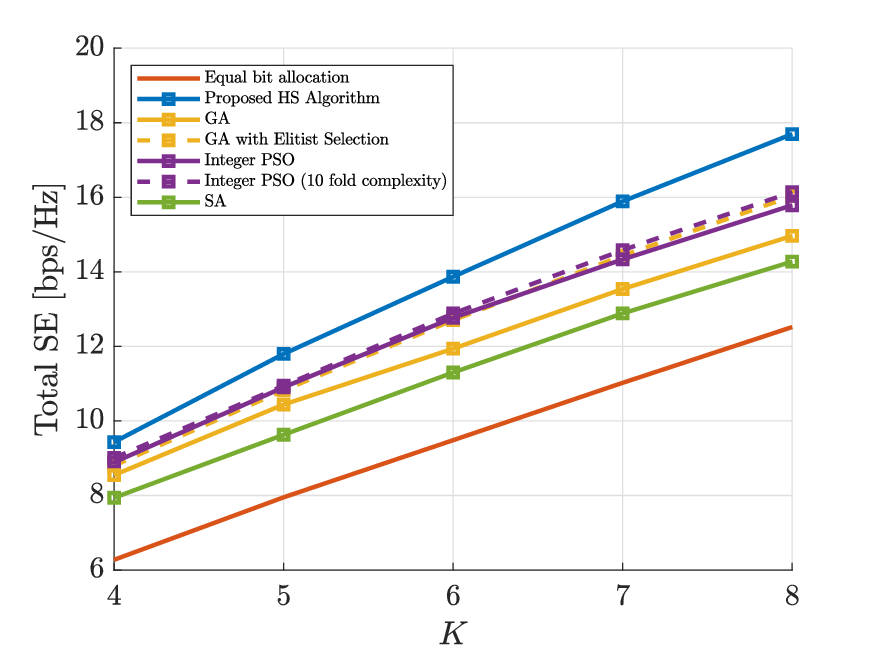}%
	\caption{The total SE performance comparison between the proposed HS algorithm, GA, integer PSO, and SA. }
	\label{fig: sum_compare}
\end{figure}

\begin{figure}[t]
	\centering
	\includegraphics[width=9 cm]{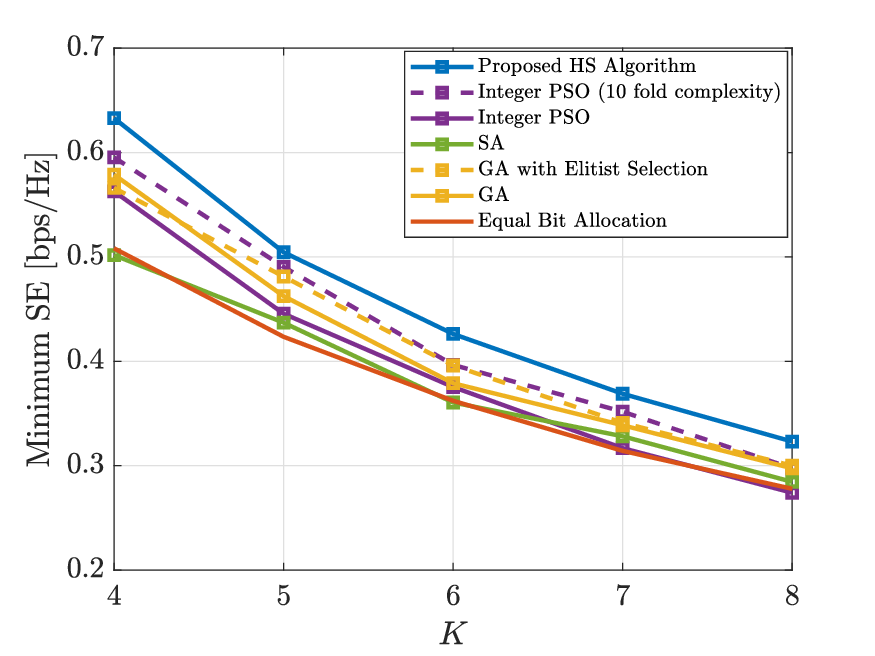}%
	\caption{The minimum SE performance comparison between the proposed HS algorithm, GA, integer PSO, and SA.}
	\label{fig: min_compre}
\end{figure}

	We investigate the effectiveness of the proposed HS algorithm by comparing it with other meta-heuristic methods such as  GA, integer PSO, and SA in Figs.~\ref{fig: sum_compare} and \ref{fig: min_compre}. 
	To ensure a balanced and thorough comparison, both stages – specifically, \textit{Stage 1} and \textit{Stage 2} – are applied for all algorithms. 
	In striving for consistency in computational complexity across these methods, we ensure the number of newly generated solutions remains the same. Additionally, we include variations like \textit{GA with Elitist Selection}\footnote{Elitist Selection is a method where the next generation of individuals is selected based on the highest fitness values \cite{ElitistSelection}.} and \textit{integer PSO (10 fold complexity)}, which exhibit increased computational complexities due to a greater number of solutions and iterations. These specific variants allow us to better illustrate the superiority of the proposed HS algorithm.
	We use an uniform crossover for GA to avoid the additional loss due to ordering. The specific parameters of algorithms are listed up in Table~\ref{size}, which includes population sizes, number of newly generated solutions per iteration, and number of iterations.
	
	According to Fig.~\ref{fig: sum_compare}, the proposed HS algorithm consistently demonstrates superior total SE performance in the fronthaul bit allocation problem when compared to  GA, integer PSO, and SA.
	A significant observation is the increasing performance gap between the proposed HS algorithm and the other algorithms as the number of UEs grows. 
	We attribute the enhanced performance of the proposed HS algorithm to its solution update process.
	Unlike GA, which relies on a pair of solutions to make a new solution and requires an additional task of determining a crossover point, the proposed HS algorithm capitalizes on the collective information of all existing solutions to find a new feasible one.
	This approach offers an advantage especially when the solution vector has a large dimension. 
	Even when we utilize Elitist Selection for GA, in which we further consider all possible parent crossover combinations to maximize the effectiveness of GA, the proposed HS algorithm still outperforms GA. Since GA relies on a set of initial parents for the generation of offspring,  inherently limiting the extent of the search space exploration, it may suffer from the proliferation of redundant solutions, known as premature convergence \cite{premature}.
	While integer PSO outperforms both GA and SA, our proposed algorithm still leads in terms of overall effectiveness. 
	The performance of integer PSO is diminished due to its discretization process, affecting the speed of particle and ability to escape the local optima. Despite increasing the computational complexity of integer PSO by tenfold with 10-times more iterations, the results consistently highlight the superior performance of the proposed HS algorithm.
	Since SA belongs to the trajectory-based algorithm that generates new solutions in an arbitrary fashion without leveraging the existing population, the performance of SA further decreases when the dimensional of the solution vector increases.
	
	In Fig.~\ref{fig: min_compre}, we compare the meta-heuristic algorithms with a minimum SE metric. Similar to the total SE result in Fig.~\ref{fig: sum_compare}, the proposed HS algorithm shows the best performance. Some variants, such as GA with Elitist Selection and integer PSO with a tenfold increase in iteration numbers, exhibit enhanced exploitation capabilities, leading to improved performance over their original forms. However, as both GA and integer PSO rely on an initial set of solutions for updates and do not generate completely new solutions, their exploration potential is limited. In contrast, the proposed HS algorithm amplifies its exploration scope through a process of random generation. Thus, the proposed HS algorithm can be superior to others in the max-min fairness scenario as well as the total SE maximization scenario. While SA employs random solutions only, its lack of a population results in an absence of directional updates and relatively low performance.

	\begin{figure}[t]
		\centering
		\includegraphics[width=9 cm]{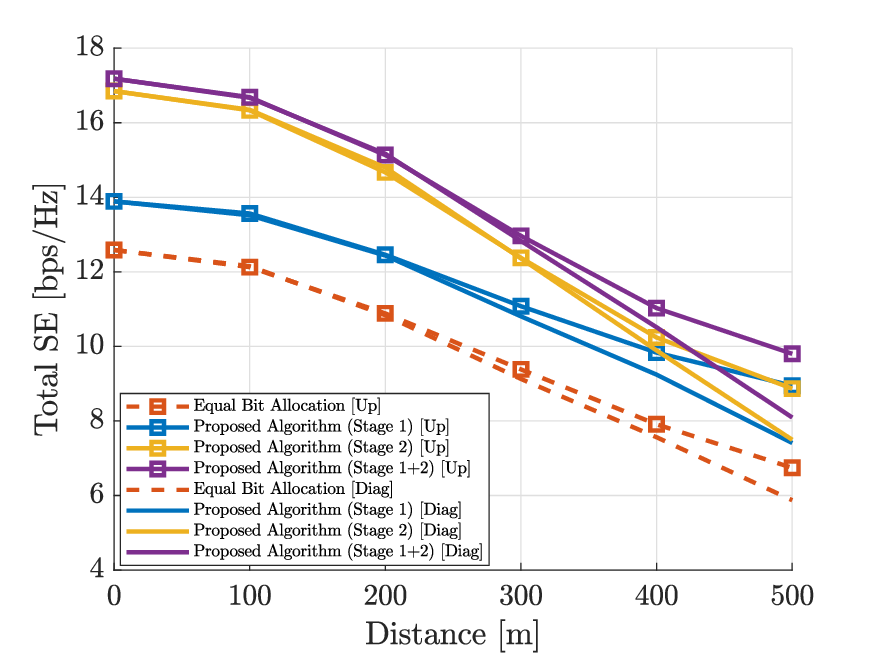}%
		\caption{The total SE versus the distances along two directions (upward and diagonal) with the wide UE distribution (${1\times1}$~$\mathrm{km}^2$). }
		\label{sum_wide}
	\end{figure}

	We plot Fig.~\ref{sum_wide} to examine the performance of the proposed algorithm in different scenarios where the center of the UE distribution is displaced from the origin by a distance ranging from 0 to 500 meters. We set two displacement directions for the simulations, upward and diagonal. Relying on Fig.~\ref{setup}, the upward direction refers to the route passing through between AP~3 and AP~4, while the diagonal direction denotes the path toward AP~4. We allow the UEs to be distributed in the wide area, $1 \times 1$ $\mathrm{km}^2$.
	According to Fig.~\ref{sum_wide}, the total SE decreases as the center of UE distribution is located far away from the origin. 
	As UEs move away from the origin, some of them experience larger pathloss, while those already in close proximity to the APs are unable to further reduce their distance, which leads to the performance degradation.
	When the center of the UE distribution is shifted upwards, there is a higher likelihood of preserving shorter distances between the APs and UEs compared to the diagonal direction scenario. Consequently, the performance exhibits a slower decline. The gap between the proposed algorithm and the equal bit allocation, however, still exists in both scenarios.

	\begin{figure}[t]
		\centering
		\includegraphics[width=9 cm]{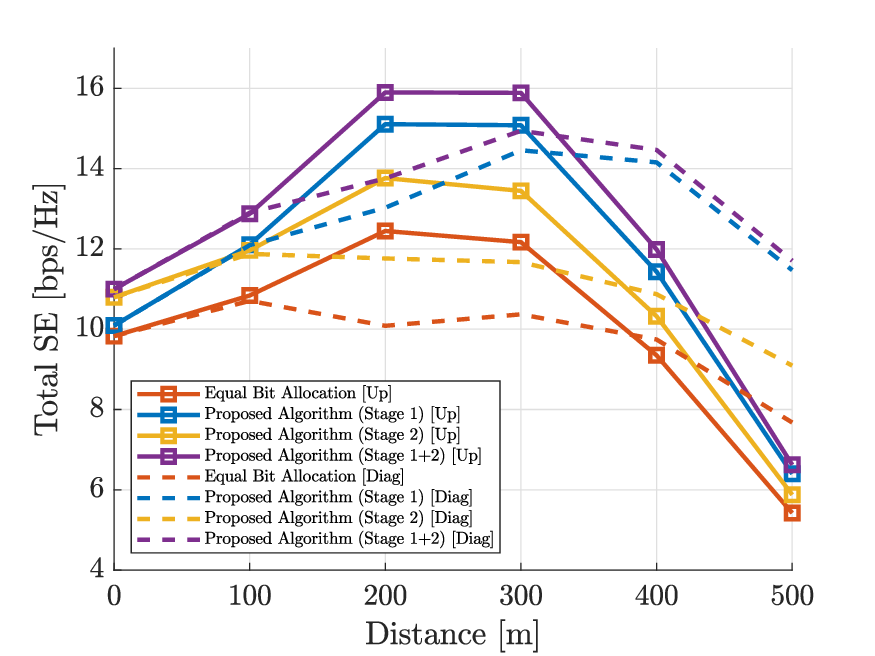}%
		\caption{The total SE versus the distances along two directions (upward and diagonal) with the dense UE distribution (${250\times250}$~$\mathrm{m}^2$).}
		\label{sum_dense}
	\end{figure}

	In Fig.~\ref{sum_dense}, we investigate the effect of fronthaul bit allocation in a highly dense UE scenario, where the simulation setups are identically set to Fig.~\ref{sum_wide} except for the area size of UE distribution, which is $250\times250$ $\mathrm{m}^2$.
	In contrast to the wide UE distribution scenario of Fig.~\ref{sum_wide},
	the UEs are located far away from the APs when the center of distribution is positioned at the origin. Moreover, each UE encounters severe inter-user interference due to the proximity of neighboring UEs, resulting in lower performances at the origin than the wide scenario in Fig.~\ref{sum_wide}. 
	
	The performances of two directions are rising until the center has the closest distance from the APs, i.e., $250$ $\mathrm{m}$ for the upward direction and $250\sqrt{2}$ $\mathrm{m}$ for the diagonal direction. This is because all UEs, including the farthest UE and the nearest UE, can experience the lower pathloss as the center approaches the APs, leading to performance improvement.  Moving along the diagonal direction is more advantageous than moving upwards, as AP~4 lies on the straight line of the diagonal path, leading to a shorter minimum distance between the nearest UE and the dominant AP when the center is located at the position of AP~4. 
	As depicted in Fig.~\ref{sum_dense}, Stage 2 seems to underperform compared to Stage 1 when the distance increases. This discrepancy can be attributed to the fact that some APs emerge as the dominant force in that scenario. While Stage 1 is designed to optimize fronthaul bit allocation among different APs, Stage 2 focuses on the bit allocation for UEs. Therefore, operating under the assumption that all APs share the same available bits may be suboptimal. It could be more advantageous to allocate a larger portion of bits to the dominant AP, thereby potentially enhancing performance.

	\begin{figure}[t]
		\centering
		\includegraphics[width=9 cm]{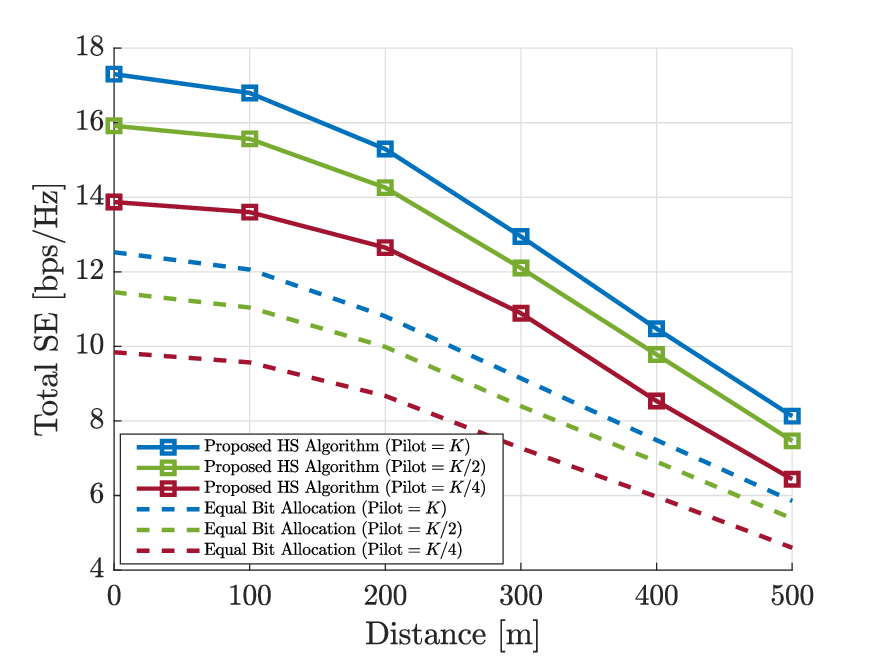}%
		\caption{The pilot contamination effect on the total SE (versus the distance along the upward direction with the wide UE distribution). }
		\label{fig: pilot}
	\end{figure}

	In Fig.~\ref{fig: pilot}, we explore the impact of pilot contamination on system performance. To  introduce non-orthogonality in a pilot sequence set, we consider shorter channel estimation time with lengths of $\tau_c = K/2$ and $\tau_c = K/4$. It is worth noting that pilot sequence assignment plays a pivotal role in overall system efficacy. For this reason, we employ the greedy pilot assignment strategy, as outlined in \cite{Ngo:2017}, instead of a random assignment approach. As expected, when all the UEs have  orthogonal pilot sequences the result shows the best performance, and the performances are reduced as the length of the pilot sequence decreases.
		Still, the efficacy of our proposed algorithm is evident as it improves SE even amidst pilot contamination.

	\begin{figure}[t]
		\centering
		\includegraphics[width=9 cm]{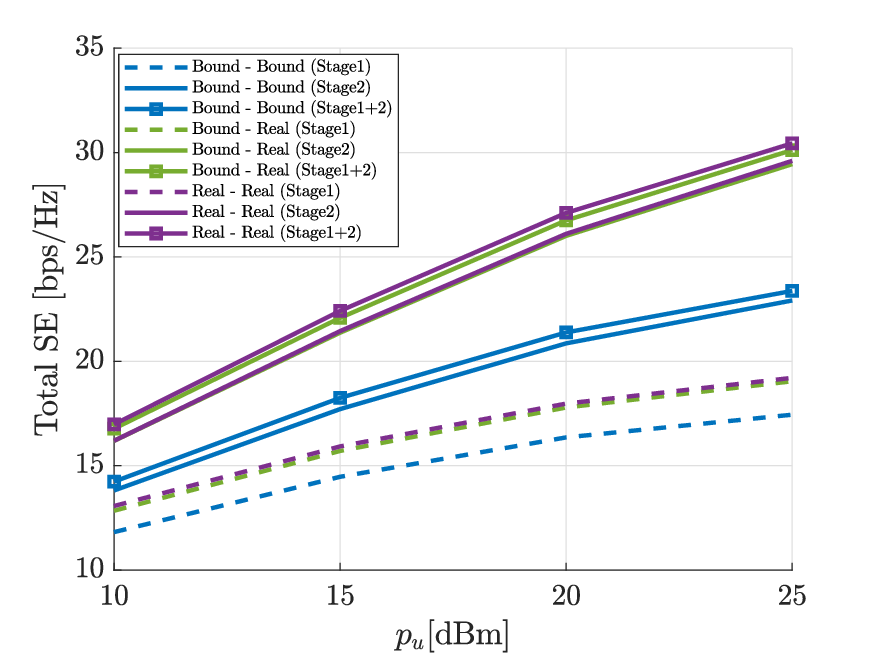}%
		\caption{The comparison between the UatF bound and the real SINR (versus the UE transmit power).}
		\label{figure7}
	\end{figure}

	\begin{figure}[t]
	\centering
	\includegraphics[width=9 cm]{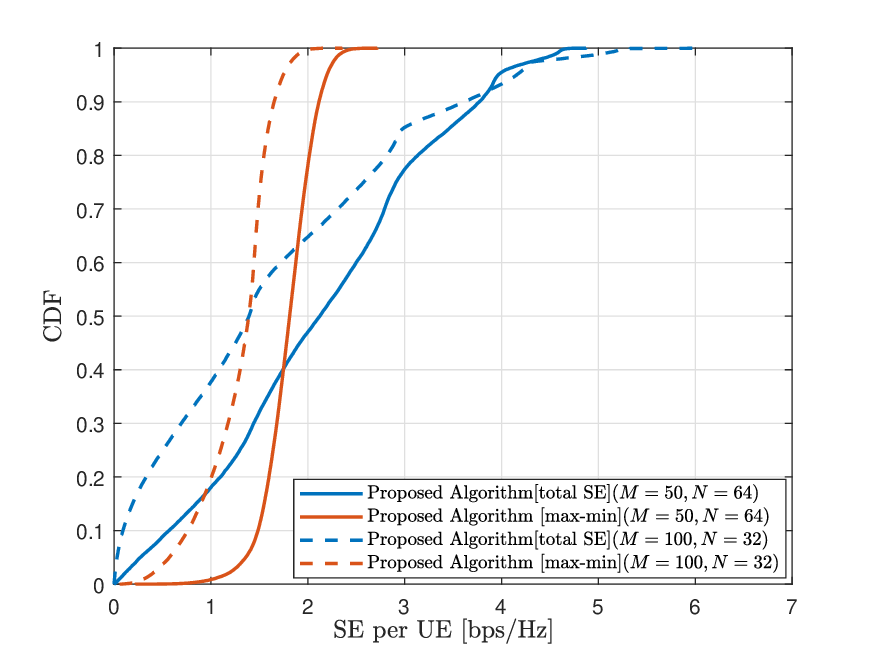}%
	\caption{The SE per UE in massive setups.}
	\label{fig: large AP2}
\end{figure}

In Fig.~\ref{figure7}, we investigate the tightness of the use-and-then-forget (UatF) bound \cite{bjornson2017} that is assumed in \eqref{SINR1}. \textit{Bound} stands for the UatF bound and \textit{Real} denotes the real SINR value, and We categorize them as \textit{Bound-Bound}, \textit{Bound-Real}, and \textit{Real-Real}, with the initial term indicating the foundational data for the bit allocation procedure and the latter signifying the SINR measure for total SE calculations.
	We evaluate the total SE concerning the UE transmit power, which varies from $10$ dBm to $25$ dBm. As the transmit power of UE increases, the performances improve while the SE will eventually saturate due to the inter-user interference. 
	According to Fig.~\ref{figure7}, the outcomes of these evaluations shows a performance gap between the UatF bound and the real SINR as already shown in \cite{bjornson2020scalable}. Nonetheless, a key observation is that the fronthaul bit allocation strategy derived from the UatF bound exhibits commendable performance with the case we take a bit allocation strategy based on the real SINR, which are denoted as \textit{Bound - Real} and \textit{Real - Real}, respectively. Therefore, despite the gap between the UatF bound and the actual SINR, the UatF bound can provide initial strategies for fronthaul bit allocation.

	Additionally, to further validate the robustness and scalability of our approach in more extensive setups, we conducted simulations for larger configurations. These experiments encompassed scenarios with increased numbers of APs, specifically ($M=100, N=32, K=20$) and ($M=50, N=64, K=20$). 
		As depicted in Fig.~\ref{fig: large AP2}, the proposed algorithm with respect to the total SE outperforms the algorithm with respect to the minimum SE for high SE per UE region. However, for UEs having lower SE, the max-min fairness proposed algorithm can provide better performance than the other. Compared to the case with larger number of APs ($M=100$), it is more efficient using many antennas in smaller number of APs ($M=50$), which shows that making favorable channel is important to enhance the system performance.

	\section{Conclusion}\label{conclusion}
	In this paper, we investigated the effect of fronthaul bit allocation in fronthaul-limited cell-free massive MIMO systems. We employed the AQNM quantizer to derive how the number of fronthaul bits affects the quantization distortion and calculated the closed-form expression of SINR. The optimization problem of maximizing the total SE was tackled, and we proposed the HS-based algorithm for solving the combinatorial non-convex problem. In the proposed algorithm, the receiver filter was optimized via the generalized Rayleigh quotient method given the fixed fronthaul bit allocation, and the fronthaul bits were updated through the hierarchical HS-based approach, which effectively allocates bits for APs and UEs, respectively.
	Also, we analyzed the proposed algorithm in terms of computational complexity and convergence. In addition, we showed that the proposed algorithm can be applied to the other objective, i.e., the max-min SE. Finally, our simulation results demonstrated that the proposed algorithm offers acceptable performance with low computational complexity and tackled the issues such as comparison between meta-heuristic algorithms, pilot contamination and UatF bound tightness.

	\section*{Appendix A \\ Closed-form Expression of SINR}
	To obtain the analytical expression for the $\mathrm{SINR}_k$ specified in \eqref{SINR1}, it is necessary to calculate the values of $\left|r_k^D\right|^2$, $\mathbb {E}  \{\left |{ r_k^B }\right | ^{2} \}$, $\mathbb {E}  \{ \left |{ r_{kk'}^I} \right | ^{2} \}$, $\mathbb {E} \{\left |{ r_k^N } \right| ^{2} \}$ and $\mathbb {E}  \{ | { r_k^Q } | ^{2} \}$.
	From the results of \cite{Ngo:2017} where the fronthaul bit allocation is not considered, we can obtain
	\begin{align}
		r_k^D &= \mathbb{E} \left [ \sum_{m=1}^M
		u_{mk}(1-\rho_{mk})\sqrt{p_u} \hat{\bg}_{mk}^\mathrm{H} {\bg}_{mk}   \right ] \notag
		\\ &= N\sqrt{p_u} {\sum _{m=1}^{M}u_{mk}(1-\rho_{mk}) \gamma_{mk}},
	\end{align}
	where $(1-\rho_{mk})$ ensures the distinct quantization distortion for each link between the CPU and APs. Note that this formulation also differs to the derivations in \cite{Bashar:2019:energyefficiency} since those consider the equal bit allocation.
	Thus, the desired signal term is given by 
	\begin{align}
		\left|r_k^D\right|^2 &= p_u(N{\sum _{m=1}^{M}u_{mk}(1-\rho_{mk}) \gamma_{mk}})^{2} \notag
		\\ & = p_u N^2 \bu_k^\mathrm{H}(\bOmega_k^\mathrm{H}\bgamma_k\bgamma_k^\mathrm{H}\bOmega_k)\bu_k. \label{rD}
	\end{align}
	The beamforming uncertainty term $|r_k^B|^2$ also can be computed by
	\begin{align}
		\mathbb {E} \left \{\left |{ r_k^B  }\right | ^{2}\right \} &= \mathbb {E} \Biggl \{ \sqrt{p_u} \Biggl ( \sum_{m=1}^M
		\tilde{u}_{mk} \hat{\bg}_{mk}^\mathrm{H} {\bg}_{mk}  \notag
		\\ & ~~~~~~~~~~~~~~~~ -\mathbb{E} \left [ \sum_{m=1}^M \tilde{u}_{mk} \hat{\bg}_{mk}^\mathrm{H} {\bg}_{mk} \right ] \Biggr ) \Biggr \} \notag
		\\ & \stackrel{(a)}{=} p_uN\sum_{m=1}^M (u_{mk}(1-\rho_{mk}))^2 \gamma_{mk}\beta_{mk} \notag
		\\ & = p_uN \bu_k^\mathrm{H}(\bOmega_k^\mathrm{H}\bD_{kk}\bOmega_k)\bu_k, \label{rB}
	\end{align}
	where (a) comes form the result in \cite{Nayebi:2017}. Furthermore, we derive the inter-user interference term $|{ r_{kk'}^I}| ^{2}$ as
	
	\begin{align}
		\mathbb {E} \left \{\left |{ r_{kk'}^I}\right | ^{2}\right \}
		=&~
		\mathbb {E} \Biggl \{ \left| \sqrt{p_u} \sum_{m=1}^M  \tilde{u}_{mk}\hat{\bg}_{mk}^\mathrm{H}\bg_{mk'} \right|^2 \Biggr\} \notag
		\\ 
		=&~
		p_u\sum_{m=1}^M \tilde{u}_{mk}^2 \mathbb {E} \Biggl \{ | \hat{\bg}_{mk}^\mathrm{H}(\hat{\bg}_{mk'}+\bee_{mk'}) |^2 \Biggr\} \notag
		\\
		=&~
		p_u\sum_{m=1}^M \tilde{u}_{mk}^2  ( N\gamma_{mk} \gamma_{mk'} + N\gamma_{mk}\epsilon_{mk'}  ) \notag
		\\
		=&~
		p_u\sum_{m=1}^M (u_{mk}(1-\rho_{mk}))^2 N\gamma_{mk}\beta_{mk'} \notag
		\\ =&~ p_uN \bu_{k}^\mathrm{H}(\bOmega_k^\mathrm{H}\bD_{kk'}\bOmega_k)\bu_{k},  \label{rI}
	\end{align}
	where $\epsilon_{mk'}$ denotes the variance of one element in $\bee_{mk'}$, which follows $\epsilon_{mk'} = \beta_{mk'}-\gamma_{mk'}$ from the definition as in \eqref{channel error}. In addition, we can obtain the noise term, which is given by
	\begin{align}
		\mathbb {E} \left \{\left|{ r_k^N } \right| ^{2}\right \}=~& \mathbb {E} \left \{\left |\sum_{m=1}^M \tilde{u}_{mk}\hat{\bg}_{mk}^\mathrm{H}\bn_{m} \right|^2 \right \} \notag \\ 
		=~& \sigma_n^2N \sum _{m=1}^{M}(u_{mk}(1-\rho_{mk}))^{2} \gamma_{mk} \notag
		\\ =~& \sigma_n^2 N \bu_k^\mathrm{H}(\bOmega_k^\mathrm{H}\bGamma_k\bOmega_k)\bu_k.\label{rN}
	\end{align}

	To complete the closed-form expression of $\mathrm{SINR}_k$, we compute the quantization distortion term $\mathbb {E}  \{ | { r_k^Q } | ^{2} \}$, which is given by
	\begin{align}
		\mathbb {E} \left \{|{ r_k^Q } | ^{2}\right \}
		=&~ 
		\mathbb {E}\left \{{\left |{\sum _{m=1}^{M}u_{mk} n_{mk}^q}\right |^{2}}\right \} \notag \\
		\stackrel{(a)}{=}&~ \sum _{m=1}^{M}u_{mk}^{2}\mathbb {E}\left \{{\left |{n_{mk}^q}\right |^{2}}\right \}
		\notag\\ =&~ \sum _{m=1}^{M}u_{mk}^{2} R_{n_{mk}^q}
		\notag\\ 
		\stackrel{(b)}{\approx}&~
		\sum _{m=1}^{M}u_{mk}^{2}\rho_{mk}(1-\rho_{mk})\mathbb{E} \left \{ \check{y}_{mk} \check{y}_{mk}^* \right \} 
		\notag \\
		=&~
		\sum _{m=1}^{M}u_{mk}^{2}\rho_{mk}(1-\rho_{mk})\mathbb{E} \left \{ |\check{y}_{mk}| ^2 \right \}, \label{qdlast}
	\end{align}
	where (a) follows from the fact that quantization distortion for each AP is independent as in \eqref{uncorrelated}, and (b) comes from the approximation of $R_{n,m}^q$ in \eqref{quaantization distortion}.
	To simply compute \eqref{qdlast}, $\mathbb{E} \left \{ |\check{y}_{mk}| ^2 \right \}$ can be further expanded as
	\begin{align} 
		\mathbb{E}& \left \{ \left |   
		\check{y}_{mk}
		\right |^2 \right \}
		\notag \\
		=&~
		\mathbb{E} \left \{ \left |   
		\sum_{k'=1}^K \sqrt{p_u}\hat{\bg}_{mk}^\mathrm{H}
		\bg_{mk'}s_{k'} + \hat{\bg}_{mk}^\mathrm{H}\bn_m
		\right |^2 \right \}
		\notag \\
		=&~
		\mathbb{E} \left \{ \left |   
		\sqrt{p_u}\hat{\bg}_{mk}^\mathrm{H}
		(\hat{\bg}_{mk}+\bee_{mk})s_{k} + \hat{\bg}_{mk}^\mathrm{H}\bn_m
		\right |^2 \right \}
		\notag  \\ & \enspace \enspace +
		\sum_{k'\ne k}^K\mathbb{E} \left \{ \left |   
		\sqrt{p_u}\hat{\bg}_{mk}^\mathrm{H}
		(\hat{\bg}_{mk'}+\bee_{mk'})s_{k'}
		\right |^2 \right \}
		\notag \\
		=&~
		\mathbb{E} \left \{ \left |\sqrt{p_u}   
		\hat{\bg}_{mk}^\mathrm{H}
		\hat{\bg}_{mk} \right |^2 + \left |\sqrt{p_u}\hat{\bg}_{mk}^\mathrm{H}\bee_{mk}\right |^2 + \left | \hat{\bg}_{mk}^\mathrm{H}\bn_m
		\right |^2 \right \}
		\notag \\ & \enspace \enspace +
		\sum_{k'\ne k}^K 
		\mathbb{E} \left \{ \left |\sqrt{p_u}   
		\hat{\bg}_{mk}^\mathrm{H}
		\hat{\bg}_{mk'} \right |^2 + \left |\sqrt{p_u}\hat{\bg}_{mk}^\mathrm{H}\bee_{mk'}\right |^2 \right \}
		\notag \\
		\stackrel{(a)}{=}&~
		(p_u N(N+1)\gamma_{mk}^2)+ (p_u 
		N\gamma_{mk} (\beta_{mk} - \gamma_{mk})) +
		\sigma_n^2N\gamma_{mk}
		\notag \\
		& \enspace \enspace +
		\sum_{k'\ne k}^K \left( 
		(p_u'  N\gamma_{mk}\gamma_{mk'})+ (p_u' 
		N\gamma_{mk} (\beta_{mk'} - \gamma_{mk'}))
		\right)
		\notag \\
		=&~
		\Big ( 
		p_u N^2 \gamma_{mk}^2 + p_uN\gamma_{mk}\beta_{mk} + N\gamma_{mk}
		\Big )
		\notag \\ & \enspace \enspace +
		\sum_{k'\ne k}^K
		\Big (
		p_uN\gamma_{mk}\beta_{mk'}
		\Big )
		\notag \\
		=&~
		p_u N^2 \gamma_{mk}^2 + \sum_{k'= 1}^K \Big (
		p_uN\gamma_{mk}\beta_{mk'}+  \sigma_n^2 N\gamma_{mk} 
		\Big ), \label{quant_input}
	\end{align}
	where (a) follows from ${\mathbb {E} \left [ \left |  \hat{\bg}_{mk}\right |^{4}\right ] = N(N+1)\gamma_{mk}^2}$. By substituting \eqref{quant_input} into \eqref{qdlast} and reformulating it into the matrix form, the quantization distortion term $\mathbb {E}  \{ | { r_k^Q } | ^{2} \}$ can be represented as
	\begin{align}
		\mathbb {E} \left \{|{ r_k^Q } | ^{2}\right \}
		=~& p_uN^2 \bu_k^\mathrm{H}((\bI-\bOmega_k)^\mathrm{H}\bGamma_k^2\bOmega_k)\bu_k
		\notag \\ & \enspace + \sum_{k'= 1}^K
		p_uN \bu_k^\mathrm{H}( (\bI-\bOmega_k)^\mathrm{H} \bD_{kk'}\bOmega_k)\bu_k
		\notag \\ & \enspace + 
		\sigma_n^2 N \bu_k^\mathrm{H}((\bI-\bOmega_k)^\mathrm{H}\bGamma_k \bOmega_k)\bu_k. \label{qdfinal}
	\end{align}
	Then, by substituting \eqref{rD}-\eqref{rN}, and \eqref{qdfinal} into \eqref{SINR}, we can obtain the closed-form expression of SINR in \eqref{SINR1}.

	\bstctlcite{IEEEexample:BSTcontrol}
	\bibliographystyle{IEEEtran}
	\bibliography{bit_allocation_references}

\end{document}